 \documentclass[preprint2]{aastex}

\usepackage{color}

\slugcomment{Not to appear in Nonlearned J., 45.}

\shorttitle{Characterization of a candidate site for the CTA at San Pedro M\'artir }
\shortauthors{Tovmassian et al.}

\begin{document}

\title{Astroclimatic Characterization of Vallecitos: A Candidate Site for the Cherenkov Telescope Array at San Pedro M\'artir }


\author{
Gagik Tovmassian\altaffilmark{1},
Mercedes-Stephania Hernandez\altaffilmark{1}, 
Jose Luis Ochoa\altaffilmark{1},
Jean-Pierre Ernenwein\altaffilmark{2},
Dusan Mandat\altaffilmark{3},
Miroslav  Pech\altaffilmark{3},
Ilse Plauchu Frayn\altaffilmark{1},
Enrique Colorado\altaffilmark{1},
Jose Manuel Murillo\altaffilmark{1},
Urania Cese\~na\altaffilmark{1},
Benjamin Garcia\altaffilmark{1},
William H. Lee\altaffilmark{1},
Tomasz Bulik\altaffilmark{4}, 
Markus Garczarczyk\altaffilmark{5},
Christian Fruck\altaffilmark{5},
Heide Costantini\altaffilmark{2},
Marek Cieslar\altaffilmark{4},
Taylor Aune\altaffilmark{6},
Stephane Vincent\altaffilmark{7},
John Carr\altaffilmark{2},
Natalia Serre\altaffilmark{8},
Petr Janecek\altaffilmark{9}
Dennis Haefner\altaffilmark{10}
}

\altaffiltext{1}{Instituto de Astronom\'ia, UNAM, Ensenada, M\'exico;} 
\email{gag@astro.unam.mx}
\altaffiltext{2}{Aix Marseille Universit\'e, CNRS/IN2P3, CPPM UMR 7346, 13288 Marseille, France}
\altaffiltext{3}{RCPTM, Joint Laboratory of Optics of Palacky University and Institute of Physics CAS, Faculty of Science, Olomouc, Czech Republic}
\altaffiltext{4}{Warsaw University Astronomical Observatory, Warsaw, Poland}
\altaffiltext{5}{Max Planck Institute of Physics, Munich, Germany}
\altaffiltext{6}{Department of Physics and Astronomy, UCLA, Los Angeles, CA, USA}
\altaffiltext{7}{Deutsches Elektronen Synchrotron (DESY), Zeuthen, Germany}
\altaffiltext{8}{LSW/ZAH, U Heidelberg, Heidelberg, Germany}
\altaffiltext{9}{Institute of Physics of the Czech Academy of Sciences,Czech Republic}
\altaffiltext{10}{Mobile Rocket Base, German Aerospace Center (DLR), D-82234 Wessling, Germany}

\begin{abstract}
We conducted an 18 month long study  of the weather conditions of the Vallecitos, a proposed site in M\'exico to harbor the northern array of the Cherenkov Telescope Array (CTA). It is located in Sierra de San Pedro M\'artir (SPM) a few kilometers away from Observatorio Astron\'omico Nacional. The study is based on  data collected by the ATMOSCOPE, a multi-sensor instrument measuring the weather and sky conditions, which was commissioned and built by the CTA Consortium.  Additionally, we compare the weather conditions of the optical observatory  at SPM to the Vallecitos regarding temperature, humidity, and wind distributions. It appears that the excellent conditions at the  optical observatory benefit from the presence of microclimate established in  the Vallecitos. 
\end{abstract}





\section{Introduction}
\label{sec:intro}
Since the early 1970s when the Observatorio Astron\'omico Nacional (OAN) began operations, it was found that the mountainous ridge  of San Pedro M\'artir (SPM) Baja California, M\'exico, is one of the few  dark areas in the world with excellent atmospheric qualities and a large fraction of cloud-free time most suitable for astronomical research. 
The current observatory employs three optical telescopes with primary mirror diameters of 2.12, 1.5, and 0.84 m. 
A few large astronomical projects considered this place as a candidate site. 
Nowadays, it is a candidate site to house the northern part of the CTA\footnote{While this paper was in the process of publication a decision has been made to built the CTA-North at La Palma.}  \citep{2013APh....43....3A}. 
The CTA  aims to increase sensitivity by another order of magnitude over currently available instruments for deep observations around 1 TeV  and significantly boost the detection area and hence detection rates at the highest energies. It intends to improve the angular resolution and hence the ability to resolve the morphology of extended sources,  to provide uniform energy coverage for photons from some tens of GeV to beyond 100 TeV, and  to enhance the sky survey capability. 
To view the whole sky, two CTA sites are foreseen. One is planned for  the southern hemisphere with a better view of  the central region of our Galaxy. A second, northern site will be primarily devoted to the study of extragalactic sources like active galactic nuclei  and will help to shape our comprehension of cosmological processes. Several candidate sites compete for hosting CTA southern and northern arrays; among them, SPM was proposed by the astronomical community of   M\'exico. 

In the  process of monitoring  the conditions of the observatory and exploration of the site for future telescopes, several   studies have been conducted with the first publication starting in 1977   \citep{1977RMxAA...2...43A,1992RMxAA..24..179T,1996AAS...188.5401C,1997RMxAA..33...59H,1998RMxAA..34...47E,2001RMxAA..37..187S,2002A&A...396..723C,2002JAD.....8....2S} to the entire volume 19 of {\sl Revista Mexicana of Astronomy and Astrophysics} dedicated to SPM  in 2003 \citep{2003RMxAC..19.....C}, and a number of recent  publications  (\citealp{2004PASP..116..682A,2006PASP..118..503A,2007RMxAC..31...71A,2011RMxAA..47...75A,2005PASP..117..104C,2007RMxAC..28....1W,2007RMxAC..31...47T,2012MNRAS.426..635S,2007RMxAC..31...93S,2007RMxAC..31..113A,2008RMxAA..44..231B,2009RMxAA..45..161O,2011RMxAA..47..409A,2012MNRAS.420.1273C}).  These studies partially dwell on  the atmospheric conditions like the turbulence, seeing, and characteristics important for operation of optical-infrared telescopes. However,  the mode of observations of atmospheric Cherenkov telescopes (ACTs)  is different. Their goal is the detection of showers comprised  of charged particles created in the Earth's  atmosphere by the  interaction of a very high energy (VHE) $\gamma$-ray with the atmosphere. Such showers induce a faint blue glow caused by the Cherenkov radiation. Hence, the astro-seeing is not a concern, but other environmental parameters are important, because a big number of  ACTs require a large flat surface to be installed and they are not protected by domes or other structures.

Here, we present results of a study of the proposed site at the SPM National Park.  The site is located  within the  area designated for the astronomical use  by a decree published in  the Official Gazette of the Federation on 2009 December 15.  The CTA Consortium has developed and constructed autonomous monitoring stations  called Autonomous Tool for Measuring Site COnditions PrEcisely \citep[ATMOSCOPE;][]{2015JInst..10P4012F} to measure  weather parameters (temperature, humidity, wind speed and direction, cloud altitude, atmospheric pressure) continuously, as well as the sky's darkness. These devices were  distributed among candidate sites for collection of homogeneous  and calibrated data. Later, they were upgraded with all-sky cameras  \citep[ASC,][]{2013arXiv1307.3880M} in order to estimate the fraction of cloud-free sky at the explored sites. We summarize here the data collected by the ATMOSCOPE at Vallecitos and compare it to the data collected by the weather station at the observatory  \citep{2003RMxAC..19...99M}. The study spans the time period of 18 months from  2012 September  to 2014 March.  CTA has technical requirements for the telescopes, including environmental parameters in which the observatory can operate. In this paper, we use them as guidance to relate measured values to the technical viability of the project. 


\section {Essentials}
\subsection{Vallecitos Site}
\label{sec:site}
\begin{figure*}[!t]\centering
   \includegraphics[width=13cm, clip]{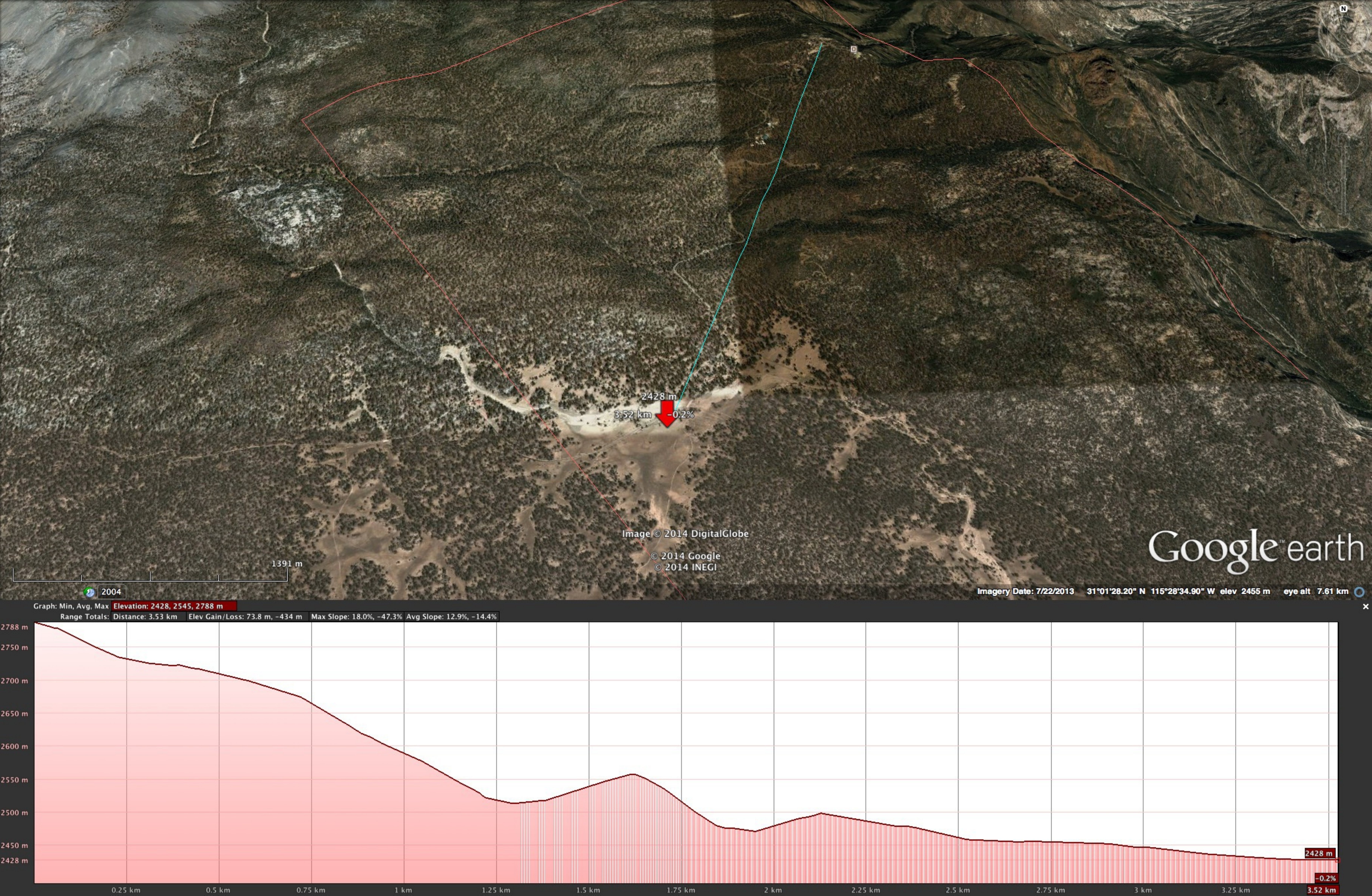}%
   \caption{Red arrow indicates the location of the ATMOSCOPE. The brownish treeless area around the red arrow is the place called {\it Vallecitos} and it has been proposed for the placement of the CTA-north array of ACTs. The light blue line indicates the distance to the weather station of the OAN SPM installed in between the 1.5 and 0.84 m telescopes. The red line indicates a rectangular area inside the SPM National Park, designated for astronomical use.}
   \label{fig:area}
\end{figure*}

The site known as {\it Vallecitos}  is located at the National Park of Sierra de SPM  Baja California, M\'exico. It is at the northern, central  part of the peninsula, which is a deserted, sparsely populated area. The  geographic coordinates of the center of the area are 31\arcdeg00\arcmin49\farcs{}20 $N$, 115\arcdeg28\arcmin39\farcs{}07 $W$ with an altitude of 2435\,m above sea level. The entire area of 1\,km$^2$ proposed to the CTA is within 3000 hectares  reserved exclusively for the astronomical research,  according to the management program of the national park in an agreement  signed in 2009 by the  federal and the state of Baja California governments.  The center of the area  is located  $\approx3500$\,m away and $370$\,m below  the weather station of the optical observatory.  The latter is located at  31\arcdeg02\arcmin40\farcs{}55 $N$, 115\arcdeg27\arcmin53\farcs{}65 $W$ at an altitude of 2802\,m above sea level. In the Figure\,\ref{fig:area} a satellite image from the Google Earth\footnote{\url{http://www.google.com/earth/}} is presented, showing the location of the ATMOSCOPE near the  area and a  line connecting the two weather stations that we use for the analysis in this paper.  Hereafter, we will refer to these two sites as  Vallecitos and OAN or observatory, respectively.

\subsection{Equipment}
\label{sec:atmoscope}

\begin{figure}[!t]\centering
   \includegraphics[width=6cm, bb = 30 30 420 800, clip]{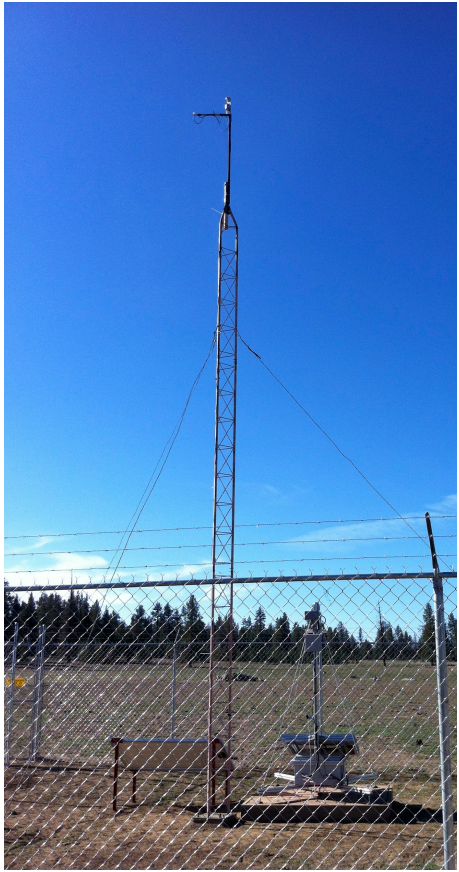}%
   \caption{The  ATMOSCOPE fully equipped and installed at the indicated location. }
   \label{fig:atm}
\end{figure}

The ATMOSCOPE  was installed  in situ (Figure\,\ref{fig:atm}) on 2012 September 21. Before that it operated for 3 months near the residential area of the observatory, while we were waiting for permission from the Environmental Protection Agency.  The ATMOSCOPE  includes a commercial weather station  from REINHARDT systems und Messelectronic GmbH (Model MWS 4M),\footnote{Manual, \url{http://www.reinhardt-testsystem.de/\_pdf/english/mws4m\_e.pdf}} which is mounted on a 10\,m tall tower above the ground. The equipment is powered by solar panels and is connected via a microwave link to the local net of the OAN, transmitting data online to the  interested parties. Apart from the weather station, as was mentioned above, it has light-of-night-sky sensor (LoNS) constructed by CTA \citep{2015JInst..10P4012F} and  the ASC \citep{2013arXiv1307.3880M}. Later (on 2012 November 12), a commercial SQM (Unihedron SQM-LE)\footnote{http://unihedron.com/projects/sqm-le/} was added to the equipment. Also, to study the dependence of temperature and humidity on the height above the ground, simple commercial temperature and humidity sensors\footnote{\url{http://www.ibuttonlink.com/products/ms-th}} were installed at  1, 4, and, 7 m altitudes.   

On the other hand, the weather station of OAN SPM was permanently installed since  the end of 1998  for continuous weather control and online reports. The meteorological station is mounted on a 6 m  tower above the ground. The controlling console of the station is placed in the observing room of the 1.5 m telescope. The station is produced by Davis (Model Vantage Pro2 Plus); all sensors are certified by the National Institute for Standard and Technology (NIST, USA). The station undergoes yearly calibrations with NIST. The online data  are available online at \url{http://www.astrossp.unam.mx}.

\section{Weather  and Sky Conditions}
\label{sec:wc}
\subsection{Fraction of clear nights}

Thanks to the presence of OAN SPM, the area has been monitored and studied for a considerably long period of time, which is very important since one or two years of observing such a complex phenomenon as climate behavior can be  misleading.  The relevant studies were conducted for the existing observatory and prospective sites of new telescopes, mostly for elevated positions around Vallecitos \citep{2008RMxAA..44..231B}.   Historical records of astronomical observations from \citet{1992RMxAA..24..179T}  show that in the period from 1982 to 1992 more than 80\% of nighttime spectroscopic observations were successfully carried out. And  about 60\% of nights had photometric quality. These results are consistent with extending the observation  to up to 20 years \citep{2003RMxAC..19...75T},  during which 80.8\%  were considered spectroscopic nights and 63.1\% photometric.  Sometimes clear-sky nights would be simply called usable instead of spectroscopic.  Assessment of photometric quality (transparency, seeing, and stability)  is based on individual observers, thus is subjective. But similar results were obtained based on a two-year satellite data analysis that was reported by \citet{2013MNRAS.429.1849C}. 

 \begin{figure*}[!t]
 \centering
   \includegraphics[width=8.9cm, bb=100 0 1600 720, clip]{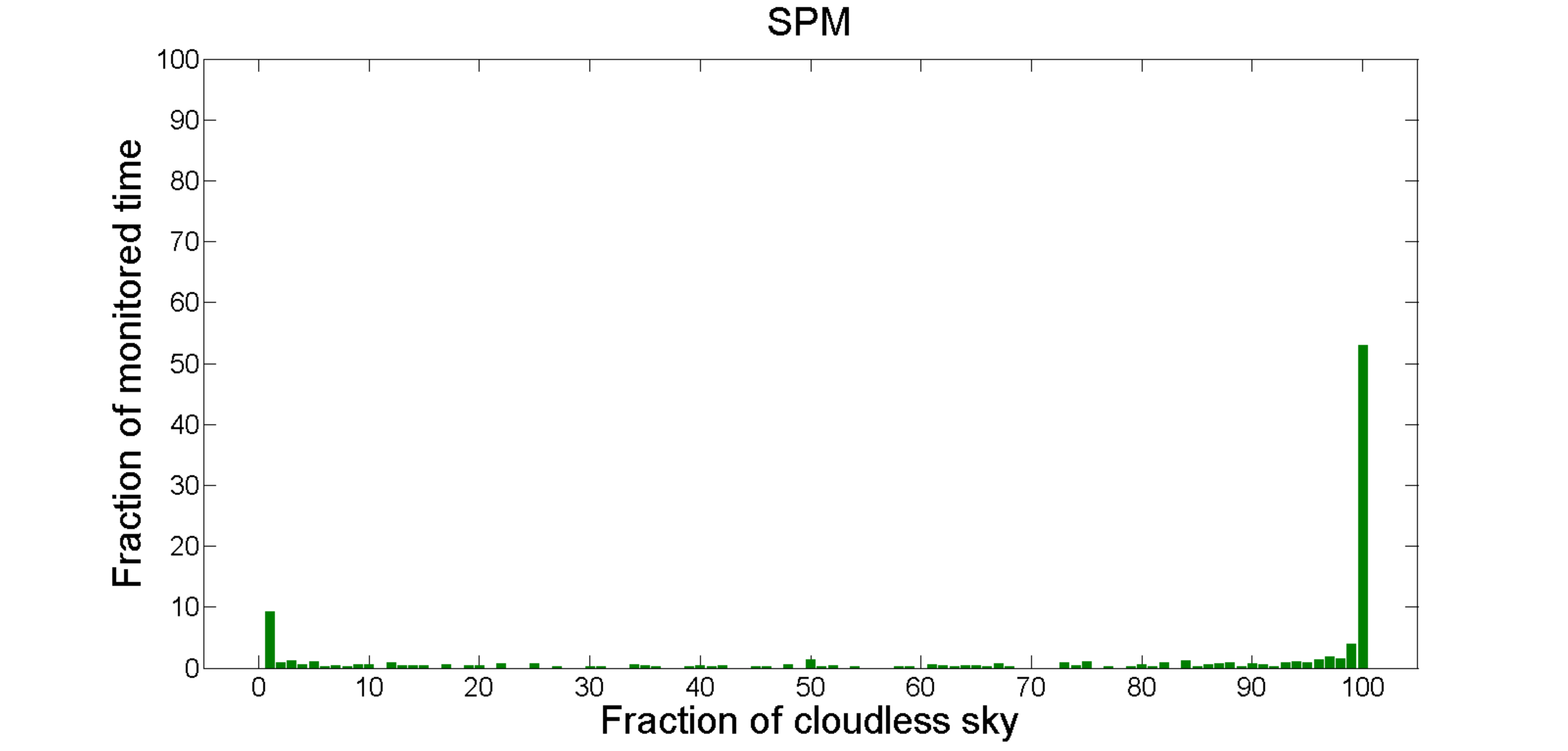}%
    \includegraphics[width=8.5cm, bb=160 0 1600 720, clip]{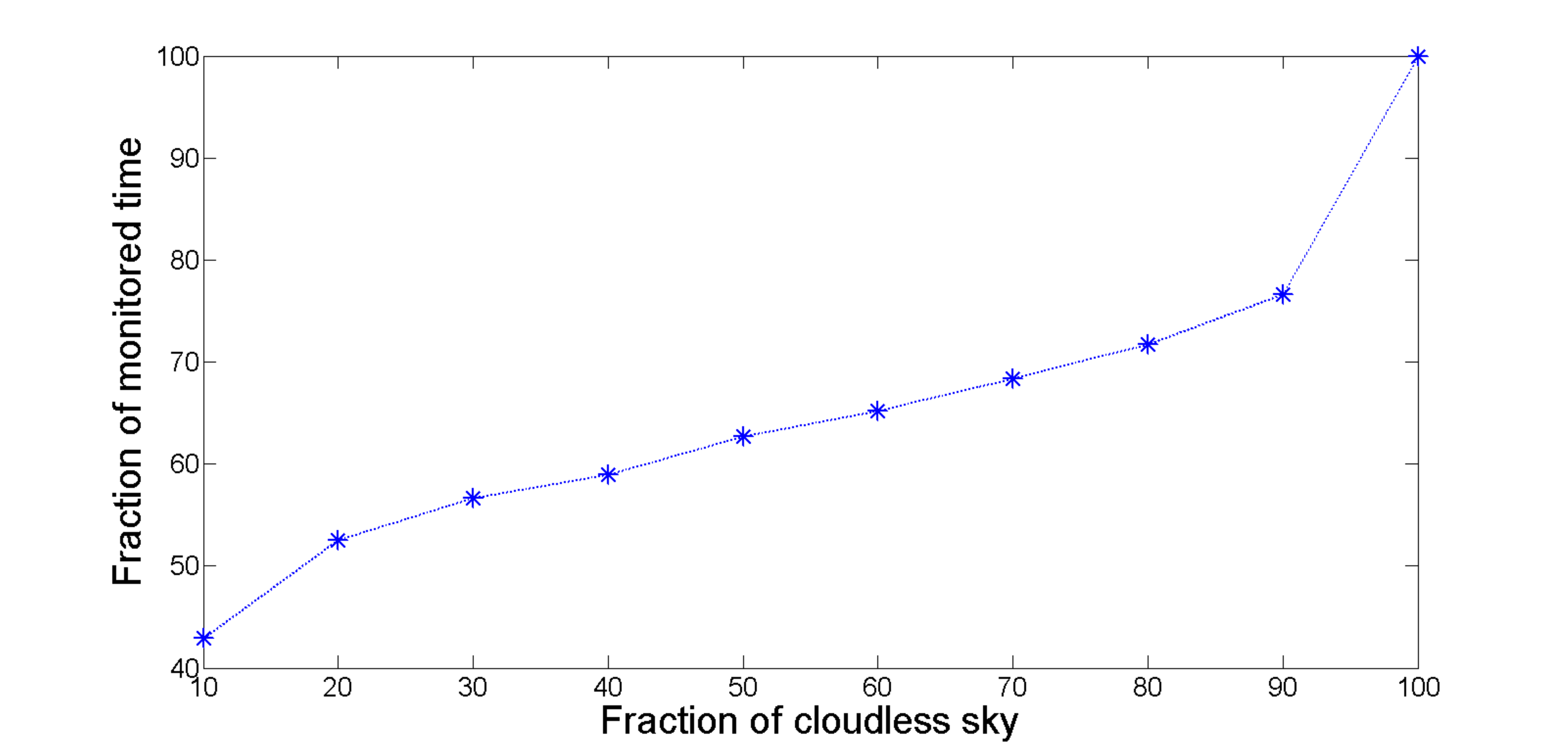}%
   \caption{ Left: distribution of cloud coverage per night during the monitored time.  Right: cumulative distribution of  cloud-free sky of Vallecitos site. The value varies from 0\% - 100\% (100, clear sky; 0, fully covered). }
   \label{fig:sky}
\end{figure*}

 Nevertheless, the definition of cloud-free conditions varies from study to study and depends on the manner of collecting data. In order to assess  all competing sites on the basis of uniform data sets, the CTA has conducted its own investigation by employing ASCs, which would determine the fraction of cloud-free time based on a star count \citep{2013arXiv1307.3880M}. Here, we present the concluding statistics. By 2014 March,  the ASC has operated 450 nights (95\% of time since the installation) taking sky images every 5 minutes. More than 18,000 images have been analyzed showing that at the  Vallecitos site $84\pm1$\% of the time $>80\%$  of the sky  is cloud free\footnote{Only moonless (the moon below horizon, so the ambient light does not affect the cloud analysis) time was taken into consideration when analyzing the ASC data. By  2014 June the number of images has increased to 20\,000 maintaining 84\% of the moonless time cloud free.}  (see Figure\,\ref{fig:sky}).  
 This result  includes evaluation of  18\% of the images from Vallecitos that had problems. They were  identified by occasions when the ASC does not see neither stars, nor the ATMOSCOPE tower located nearby. After elaborate analysis, by cross-correlating data from different  sources, including ASC images from the OAN, it was determined that part of the problematic images were the consequence of fresh fallen snow covering the lens, but the majority were affected by condensation due to the high humidity dominating the valley in contrast to the surrounding hills.  
The sky cloud coverage at the time of the problematic images was obtained by analyzing contemporary satellites images.  The correction was based on the relevant images from satellites covering the site within the analyzed time window. The data from ASC were compared with corresponding data sets from GOES satellites. The satellite images were not always available in the corresponding time window, and thus only 55\% of affected images (10\% of all data) was substituted by the satellite images. The remaining 8\% of inconclusive images was not taken into the evaluation of the fraction of clear nights of the Vallecitos site. The final result on a number of clear nights has not changed by this inclusion of the satellite images.  More details of ASC data and its analysis will be published by  \citet{in preparation}.
This is the first ground-based study in  situ based on continuous unbiased automatic measurements.   Results of the two-year study by CTA tally well with previously published statistics regarding SPM and are within year-to-year uncertainty.

\subsection{Night Sky Background}

  \begin{figure}[!t]\centering
   \includegraphics[width=7cm,height=7cm]{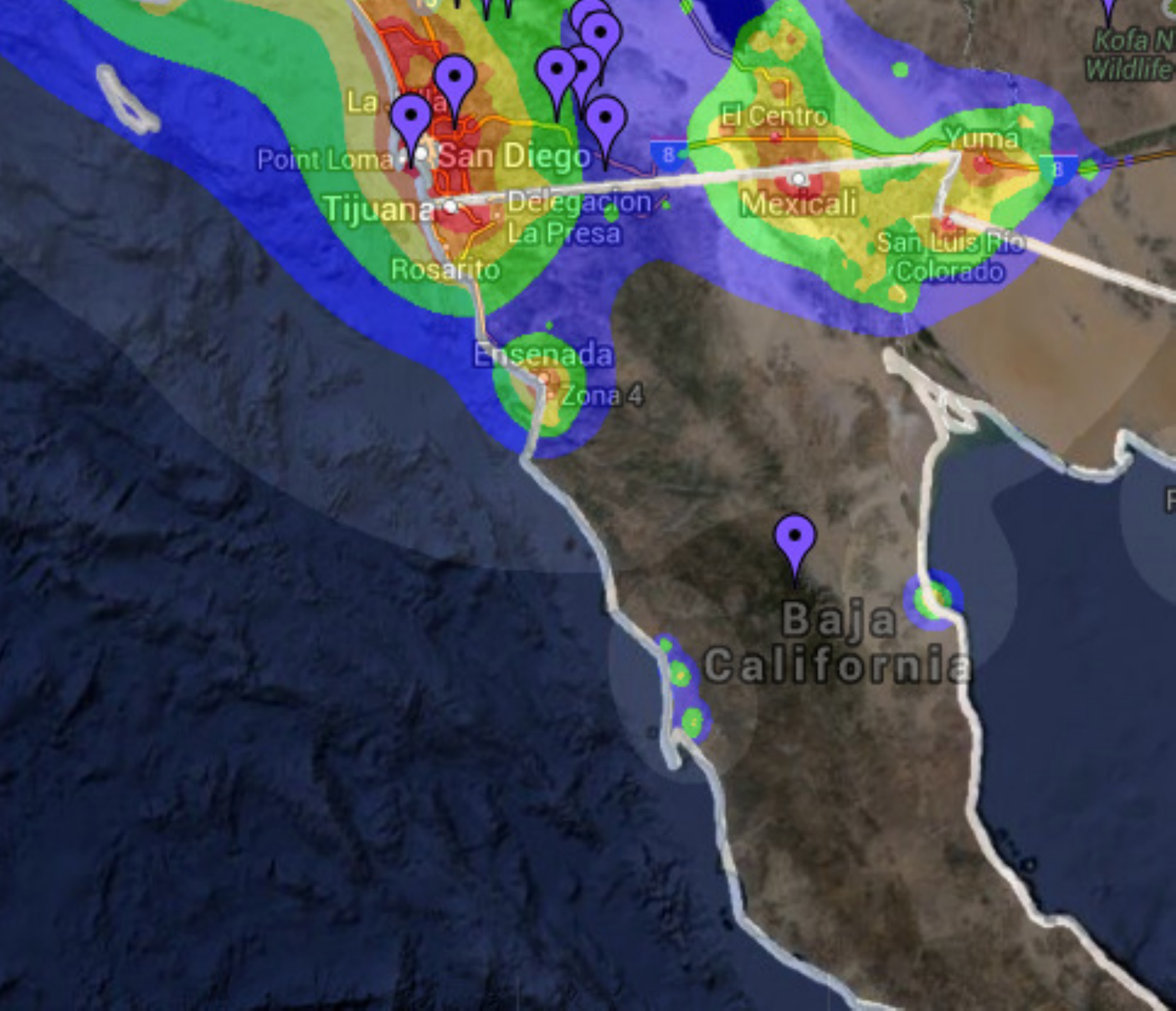}%
   \caption{ Light pollution map  of northern Baja California. The pop-up balloon at the center of  the image indicates the location of SPM and the patched area, the zone which is  affected by the nearest cities.  Source: www.jshine.net/astronomy/dark\_sky/}
   \label{fig:nsky}
\end{figure}

\begin{figure*}[!t]\centering
   \includegraphics[width=14cm]{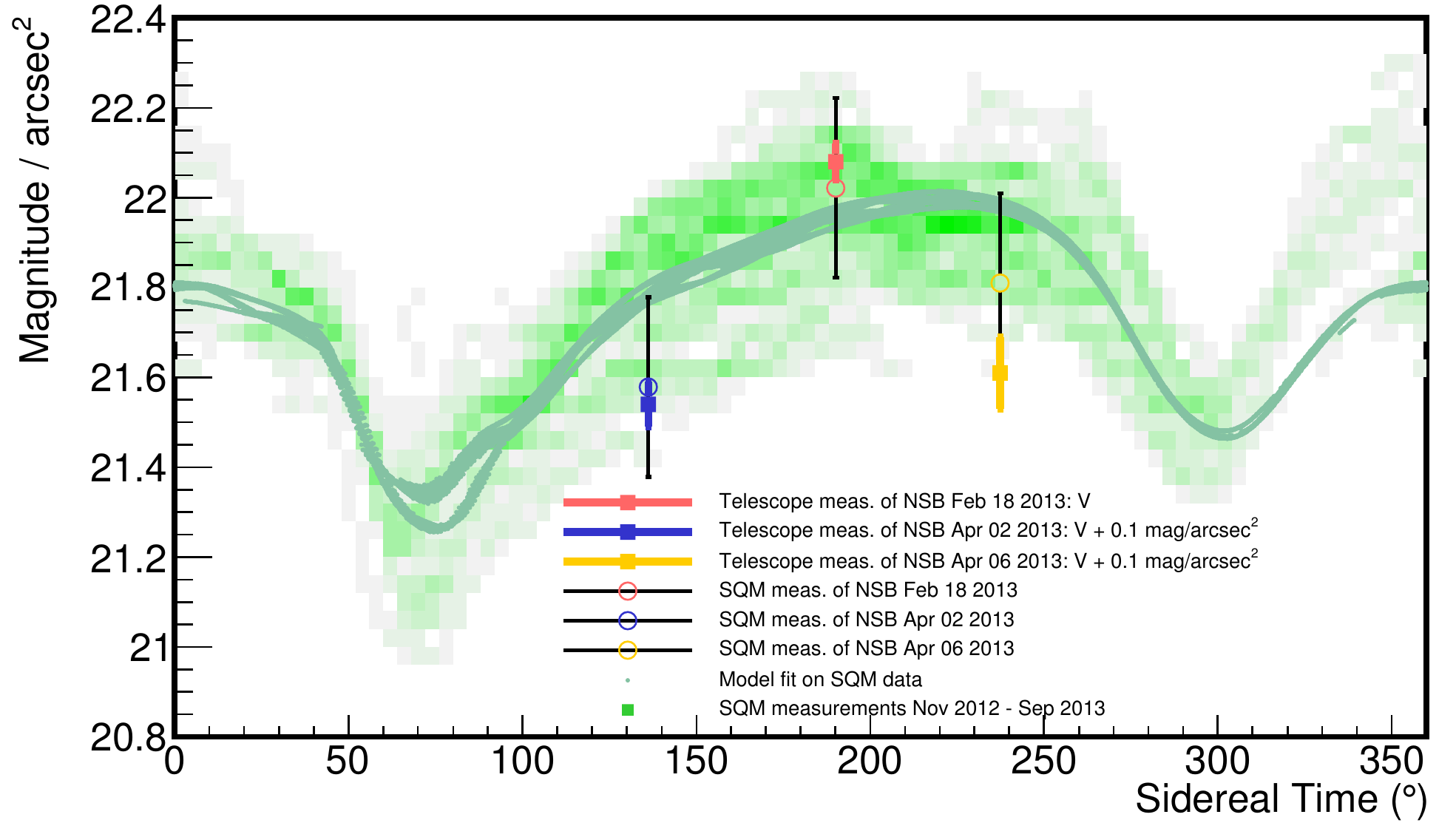}
   \caption{SQM measurements w.r.t. Sidereal Time for 10 months of data taking (area made of levels of green):
   the intensity of the green color reflects the density of points. Due to the incomplete year, the high values of S.T. 
   are less populated. A simple model (dark green curve; see the text) allows us to check the SQM behavior. 
   The spread of the SQM measurements around the model originates from the typical repeatability of an SQM (lower than $\pm$ 0.1 mag$\,$arcsec$^{-2}$), the variability of
   the atmospheric transparency, the variability of nightglow, and from possible deposits of dust on the SQM window.
   The measures of table \ref{tab:nsb},
   in V-band shifted to match the SQM band, are plotted and compared to the contemporaneous SQM measures.
   }
   \label{fig:sqm_meas}
\end{figure*}

Observatory at SPM is known for its  dark skies since it is located in a sparsely populated area.
Nevertheless,   an illumination  produced by the  urban areas along the U.S.-Mexican border and coast of the Gulf of California is notable at the horizon from the northeastern edge of the mountains where the OAN is located. Meanwhile, from the  Vallecitos site, shielded by the hills,  not a single artificial light is visible (see Figure\,\ref{fig:nsky} for the light contamination map of the area). The multi-band NSB measurements for OAN have been published by \citet{2007RMxAC..28....9T}.  In early 2013, we performed new
measurements of sky brightness to provide updated
data for the CTA.  During three separate nights, images of a Landolt standard star were obtained  along with images of the sky by  pointing to the zenith. 
The images were obtained in the U, B, V, R, and I filters with average integration times for sky of  15, 13, 
12, 9, and 8\,minutes, respectively (one image per filter). 
The first two sets (2013 February 17 and 2013 April 1) were obtained at the 0.84\,m telescope 
with  Wheel Mexman + CCD Esopo, while the last (2013 April 5) was obtained at the 
2.1m telescope with the Italian filter wheel and  CCD Marconi\,2. 
The data were processed by removal of bias and application of flat-field correction.  After preliminary reduction, an aperture photometry 
by the APT program \citep{2012PASP..124..764L} was used for the measurements. For the sky, the magnitudes were measured in 
14\arcsec\  (Esopo) and 2\arcsec\ (Marconi\,2) circular regions.
The instrumental magnitudes were corrected for atmospheric extinction using  parameters from 
\citep{2001RMxAA..37..187S}.  However, no color correction was applied for measured magnitudes. The results are presented in
 Table\,\ref{tab:nsb}.  Inspection of table values shows that  the sky was somewhat brighter in April than it was in February, which is not surprising, as there is a seasonal dependency (see Figure\,\ref{fig:sqm_meas}). 

\begin{table*}
\caption{NSB measurements at three moonless nights}
\begin{tabular}{lcccc}
\hline
\hline
Filter &            & UT Date &  & rms\\
     mag\,arcsec$^{-2} $    & 2013 Feb 18 & 2013 Apr 02 & 2013 Apr 06  & \\ 
\hline
\hline \\
U & 22.04$\pm$0.07 & 21.95$\pm$0.11 & 21.11$\pm$0.17 & 0.51 \\
B & 22.56$\pm$0.05 & 22.42$\pm$0.07 & 22.23$\pm$0.11 &  0.17 \\
V &  22.08$\pm$0.04 & 21.44$\pm$0.05 & 21.51$\pm$0.08 & 0.35 \\
R & 21.17$\pm$0.03 & 20.65$\pm$0.03 & 21.04$\pm$0.07 & 0.27 \\
I & 19.93$\pm$0.02 & 18.98$\pm$0.02 & 19.6$\pm$0.06 & 0.21 \\
\hline
\hline
\end{tabular}
\label{tab:nsb} 
\end{table*}
 The NSB evaluation by the CTA criteria is more complicated.  Identical cross-calibrated instruments were developed to perform measurements of NSB in all candidate sites in a similar, homogeneous manner. The LoNS sensor is described and data reduction and analysis methods are presented by  \citet{2013arXiv1307.3053G} and \citet{2015JInst..10P4012F}. 

 
 A commercial LoNS measurement device, a Sky Quality Meter SQM-LE, from Unihedron, is part of the ATMOSCOPE. 
 The TSL237 light-to-frequency converter, made of a silicon photodiode combined to a current-to-frequency converter, is the base of the SQM.
 The frequency measured is directly proportional to the light intensity received by the photodiode and is temperature compensated to provide
 an operating range of $\left[ -40^{\circ}{\mathrm C},\ +85^{\circ}{\mathrm C}\right]$. The measurements are then converted to  units of magnitude per square arcsecond (a common unit for surface brightness in astronomy) 
 across a field of view, which is restricted to $\sim 20^\circ$ FWHM by means of a built-in focusing lens. Our device points to the zenith.
  The spectral response of the assembly is wider than the V-band, as it also integrates  a part of the B-band. It has been shown (\citet{sqmistil}
 that this broader spectral acceptance can give an offset SQM$-$V up to 0.25 mag$\,$arcsec$^{-2}$, depending on the $B-V$ color index. In our case,
 the SQM$-$V offset is expected to be close to 0 for the observation of 2013 February 17, and close to 0.1 for the other nights from table \ref{tab:nsb}.
 Beyond the spectral systematics, a cross-calibration between several devices placed together in a controlled low-light environment has shown an intrinsic precision of the order of $\pm$ 0.2 mag$\,$arcsec$^{-2}$. 

Figure \ref{fig:sqm_meas} shows the NSB measured in the V-band (from Table\,\ref{tab:nsb}),
shifted to match the SQM band, and SQM measurements performed over 10 months: 785 hr of moonless and clear nights from early 2012 November  to early 2013 September  (1 minute periodicity).
The area made of green levels represents the data as provided by the SQM, and the curve is the expected mag$\,$arcsec$^{-2}$, from a simple model
including starlight (direct + scattered), planets, and zodiacal light, falling in the SQM field of veiw. 
This model has two  free parameters representing the light sources not taken into account (average nightglow, light pollution) and a correction
to the acceptance of the whole measurement chain (atmosphere+SQM). They are estimated using a global fit on the data. The residual NSB
extracted from the fit is 22.6$\pm$0.2 mag$\,$arcsec$^{-2}$, which makes SPM one of the darkest sites among the Northern CTA sites.
Given the global fit, the main interest of the model is to show the relevance of the variations of the SQM measures along the Sidereal Time.

Out of the SQM measures, three values are highlighted in Figure\,\ref{fig:sqm_meas}
to be compared with the NSB simultaneously measured in the V-band with the telescopes (Table\,\ref{tab:nsb}, converted in the SQM band).
The agreement is within the SQM error bars.

 
\subsection{Wind}

\begin{figure}[!t]
\centering
   \includegraphics[width=7cm,height=7cm]{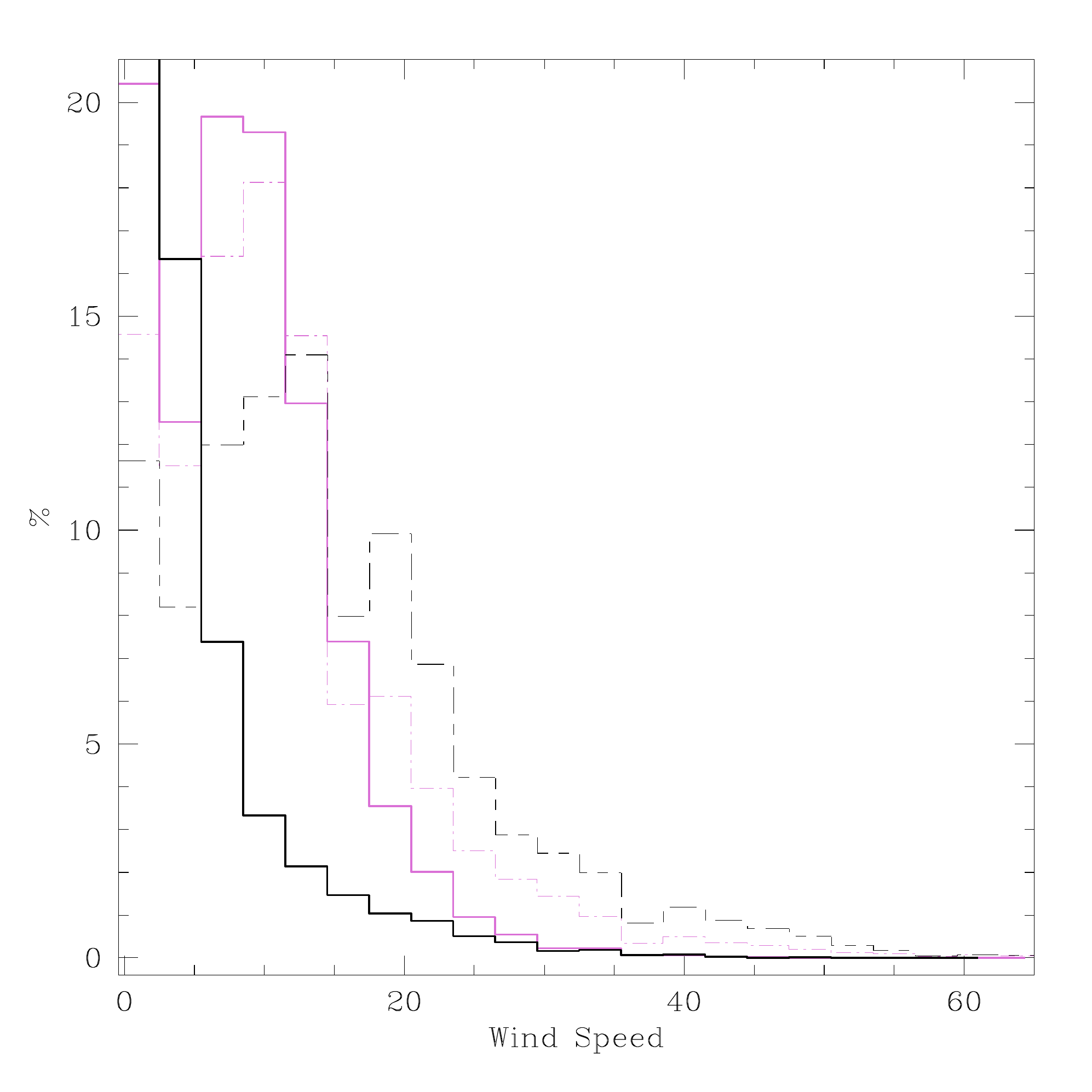}%
   \hfill
   \includegraphics[width=7cm,height=7cm]{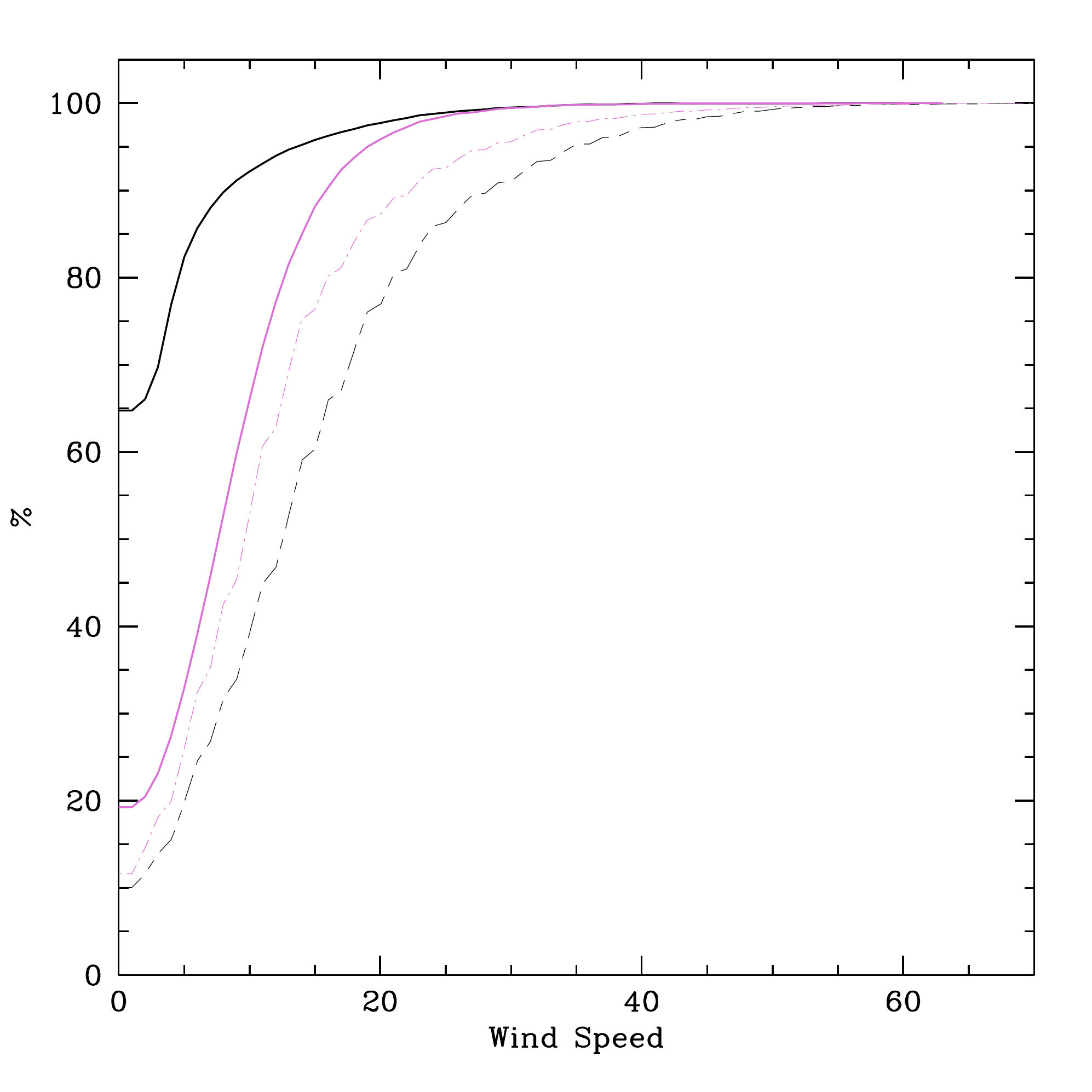}
   \caption{Top: the histogram presentation of distribution of the wind speed with 3\,km\,hr$^{-1}$ step. Bottom: the cumulative distribution function of the wind speed. In both panels,  the  Vallecitos measurements are plotted by solid lines, the black line corresponds to the nighttime (6pm to 6am local time), and the magenta line represents daytime (6am to 6pm).  Similarly, the black dashed line corresponds to SPM  night and the  magenta dot-dashed line represents the wind distribution at SPM during the day. 
}
   \label{fig:widefig1}
\end{figure}

Strong wind  or wind gusts represent a serious hazard for a telescope installed unprotected by a building and dome.   One of the constraints for  a CTA candidate site is that the wind speed does not exceed 36\,km\,hr$^{-1}$ limit in a 10 minute average during the time of observations. 
Figure \ref{fig:widefig1} presents  the distribution of wind speed in the form of a histogram on the left side and as a cumulative distribution function on the right. Since there is a notable difference between the nighttime and the daytime weather parameters, and because the nighttime measurements are more relevant, here and hereafter, the data distributions are separated between the nighttime (from 6 pm to 6 am) and the daytime, encompassing  the other half  of the day.  In addition,  data distributions  from the  Vallecitos and OAN are plotted together to allow for a direct comparison. Please note that the wind speed measurements at  Vallecitos are taken every minute, while  the OAN weather station registers it every 5 minutes   over the same period of time.  Also, that the anemometer at  Vallecitos is installed at the appropriate 10\,m height above the ground, while at the SPM it is on a 6\,m mast. However, it is feasible to  estimate the wind speed at 10\,m measured at 6\,m using a simple prescription expressed in Eq\,4 of \citet{fbretal}.  The wind measurements guidance from the federal meteorological handbook published by  Office of the Federal Coordinator for Meteorology of USA suggest to measure wind speed averaging  over two-minute period.  The wind speed data for 10 minutes shall be examined to evaluate occurrence of gusts. Gusts are indicated by rapid fluctuations in wind speed with a variation more than 18\,km\,hr$^{-1}$\  between peaks and lulls. The REINHARDT MWS 4M weather station allows averaging of instantaneous measurements by accompanying software. One-minute periodicity  reading of wind speed values measured each 2 s was programmed in our case. 

From  Figure \ref{fig:widefig1} it is obvious that  it is very quiet  during  night at the Vallecitos.  The daytime is windier, but the wind speed rarely  rises beyond  20\,km\,hr$^{-1}$. The wind speed at the OAN during  daytime is similar to the Vallecitos but generally exceeds it. At nighttime, the wind speed is statistically higher  at OAN than during the day.  In Figure \ref{fig:widefig1}, the wind speed at OAN is plotted as measured.  However, if a correction is applied for the difference of the anemometer heights, the  peak value of wind speed distribution at 14\,km\,hr$^{-1}$ in SPM at night would increase to 16 \,km\,hr$^{-1}$, shifting the entire distribution rightward. Similarly, the daytime histogram would shift too, thus slightly increasing  the gap between  Vallecitos and OAN wind speed distributions.  However, the purpose of this study is not comparison of two sites, and hence the precise corrections are irrelevant. The SPM data are brought up here only because a longer record exists for the observatory, helping project conditions at Vallecitios in the long term. 

\begin{figure}[!t]\centering
   \includegraphics[width=8cm,height=6cm]{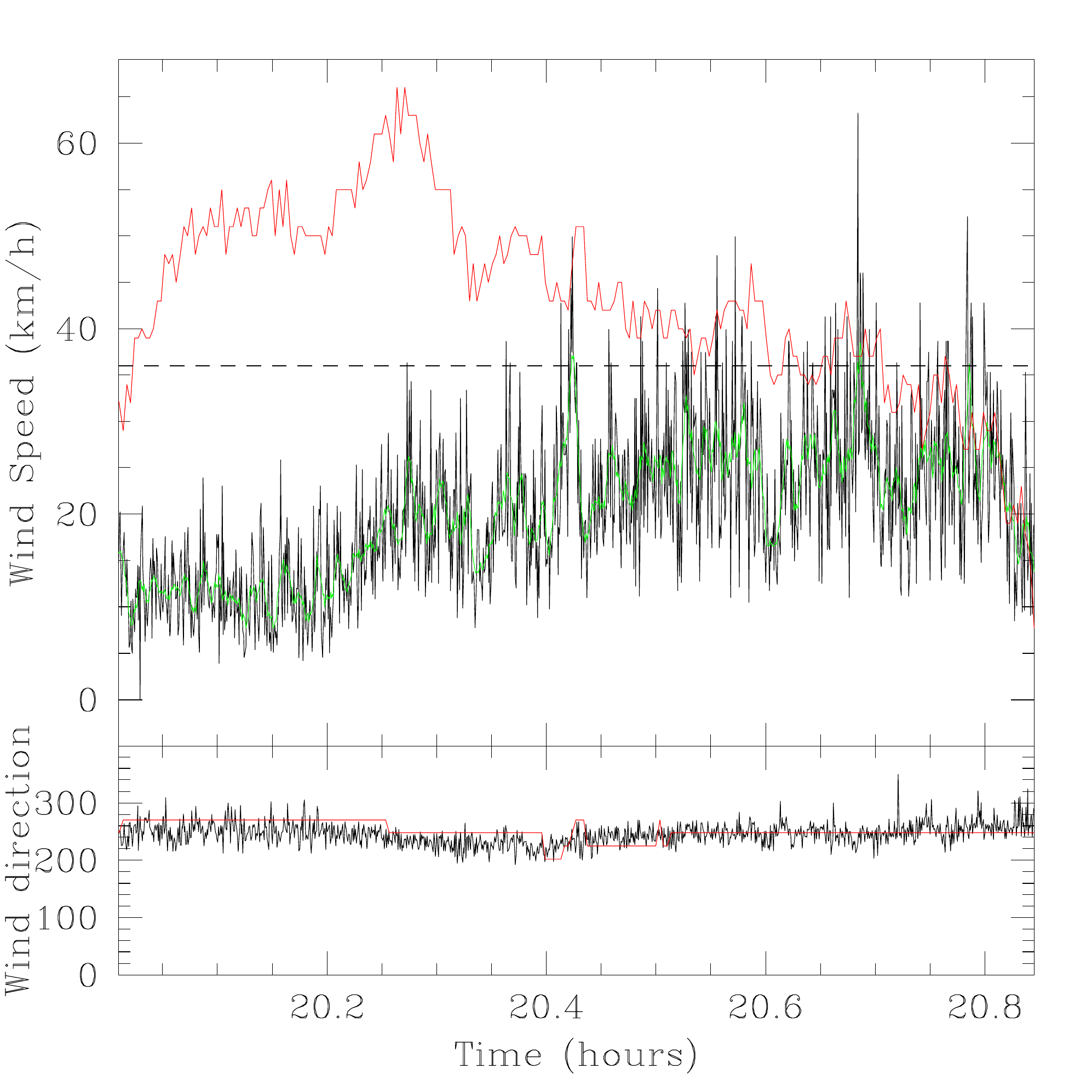}%
\caption{Example of wind speed during a storm in the area on 2013 February 20. The red line is a wind speed at SPM  in the upper panel. The black line is the wind speed at  Vallecitos .  The green line is  the Vallecitos wind speed downgraded to the same temporal resolution as SPM. The horizontal dashed line is a 36\,km\,hr$^{-1}$ operational limit. In the bottom panel, the wind direction is presented from both instruments. }
   \label{fig:gusts}
\end{figure}

Remarkably, the wind speed  at OAN does not depend much on a season according to a three-year monitoring in the 1990s \citep{1998RMxAA..34...47E}. The wind is usually  $<30$, rarely reaching 40 km\,hr$^{-1}$.  Similar results were reported by the Thirty Meter Telescope  site  exploring group, with daytime 
winds $<29 $ and nights  usually $<36$\,km\,hr$^{-1}$. 

\begin{figure}[!t]\centering
   \includegraphics[width=7.0cm,height=6cm]{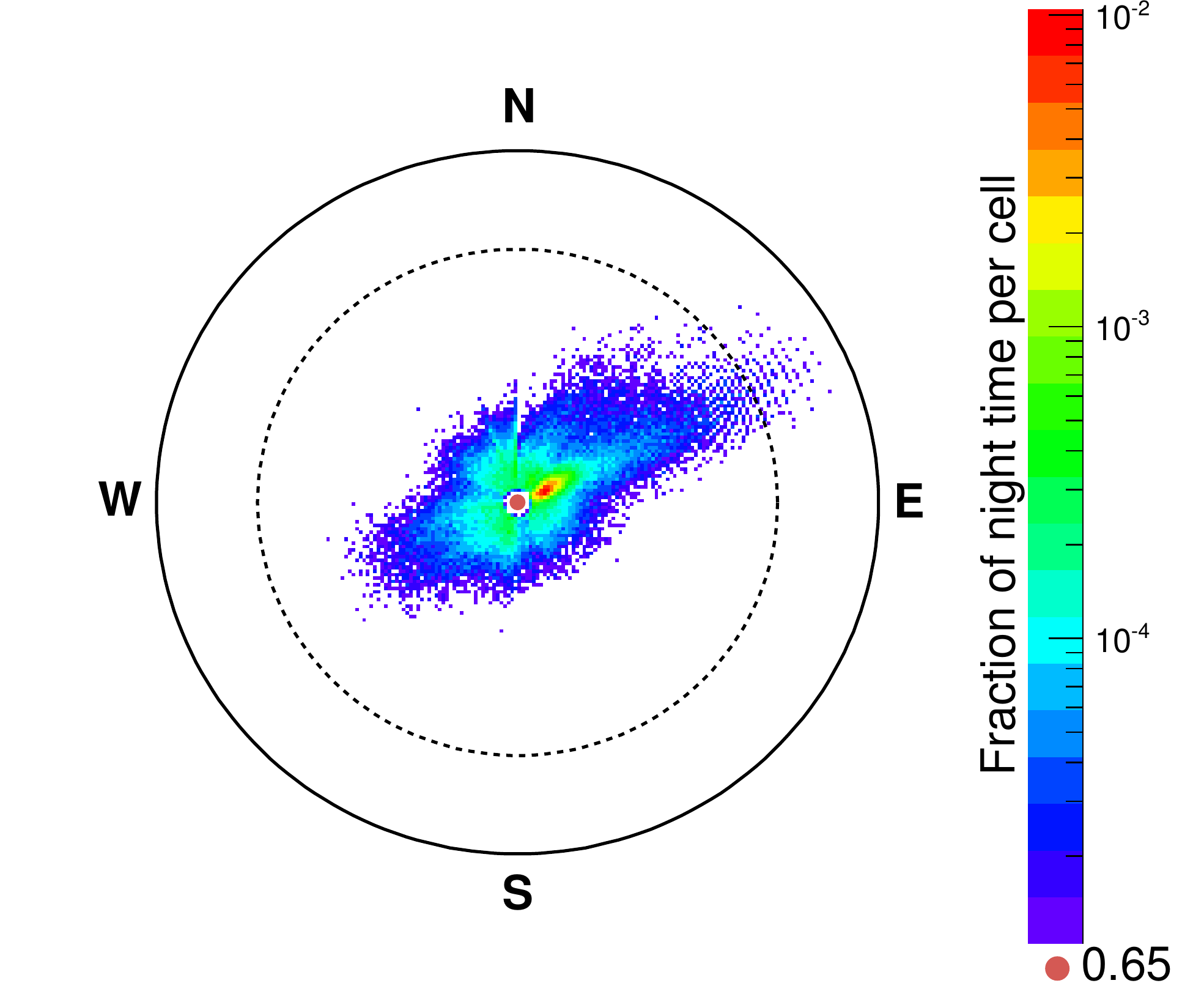}%
\caption{Wind direction distribution at  Vallecitos during night. 
The solid circle marks the 50\,km\,hr$^{-1}$  velocity and the dashed circle indicates 36\,km\,hr$^{-1}$  speed limit.  99.58\% of points lie  within dashed line circle. Brown circle in the center contains 65\% of the zero values. 
}
   \label{fig:dist_vient}
\end{figure}

The cumulative distribution function presented in the right panel demonstrates  the exceptional  tranquility of Vallecitos, where more than 65\% of the time at night   the wind speed is less than 5 km\,hr$^{-1}$, at which times anemometer cups get stuck and indicate zero value.\footnote{The anemometer starts to give values from 2\,km\,hr$^{-1}$, but  they are not very reliable, according to the specifications at speeds below 4.7\,km\,hr$^{-1}$.}  The distribution  hits the 100\% mark at low velocities,   far away from the hazard wind speed for any instrument. Meanwhile, the OAN  exhibits wind gusts of considerable speed, which is not an issue at the Vallecitos. And that is natural because the OAN station is located close to the edge of mountain ridge, while the Vallecitos is shielded by surrounding hills and trees.   In   Figure\,\ref{fig:gusts}, a wind speed record is presented on 2013 February 20,  when the region was subjected to a strong storm.\footnote{ \url{http://www.nbcsandiego.com/weather/stories/Winter-Storm-Blizzard-Snow-Hail-San-Diego-February--192023251.html}}  Evidently, at the SPM there was a strong wind reaching 70\,km\,hr$^{-1}$ around 20.2 PST. At Vallecitos it stayed calm until the storm started to wind down with only a few  gusts barely exceeding 50\,km\,hr$^{-1}$  around 20.7 PST.

 Another way to look at it  is the wind rose presented  in Figure\,\ref{fig:dist_vient}. The 
wind distribution is shown depending on the direction during the nighttime. Plots show preferential  winds in SW and NE directions. The SW direction corresponds to the orography,  a channel by which the  humid air from the Pacific enters the peninsula. The NE direction is probably related to Santa Ana winds bringing hot air from deserts, which are not frequent (a few times a year), but are rather strong.  The number of points in the ring between between 36 and 50\,km\,hr$^{-1}$ may seem significant,  but, in fact, the fraction of time when the wind exceeds 36\,km\,hr$^{-1}$ is only 0.42\% of the time.

\subsection{Temperature}

\begin{figure}[!t]\centering
   \includegraphics[width=7cm,height=7cm]{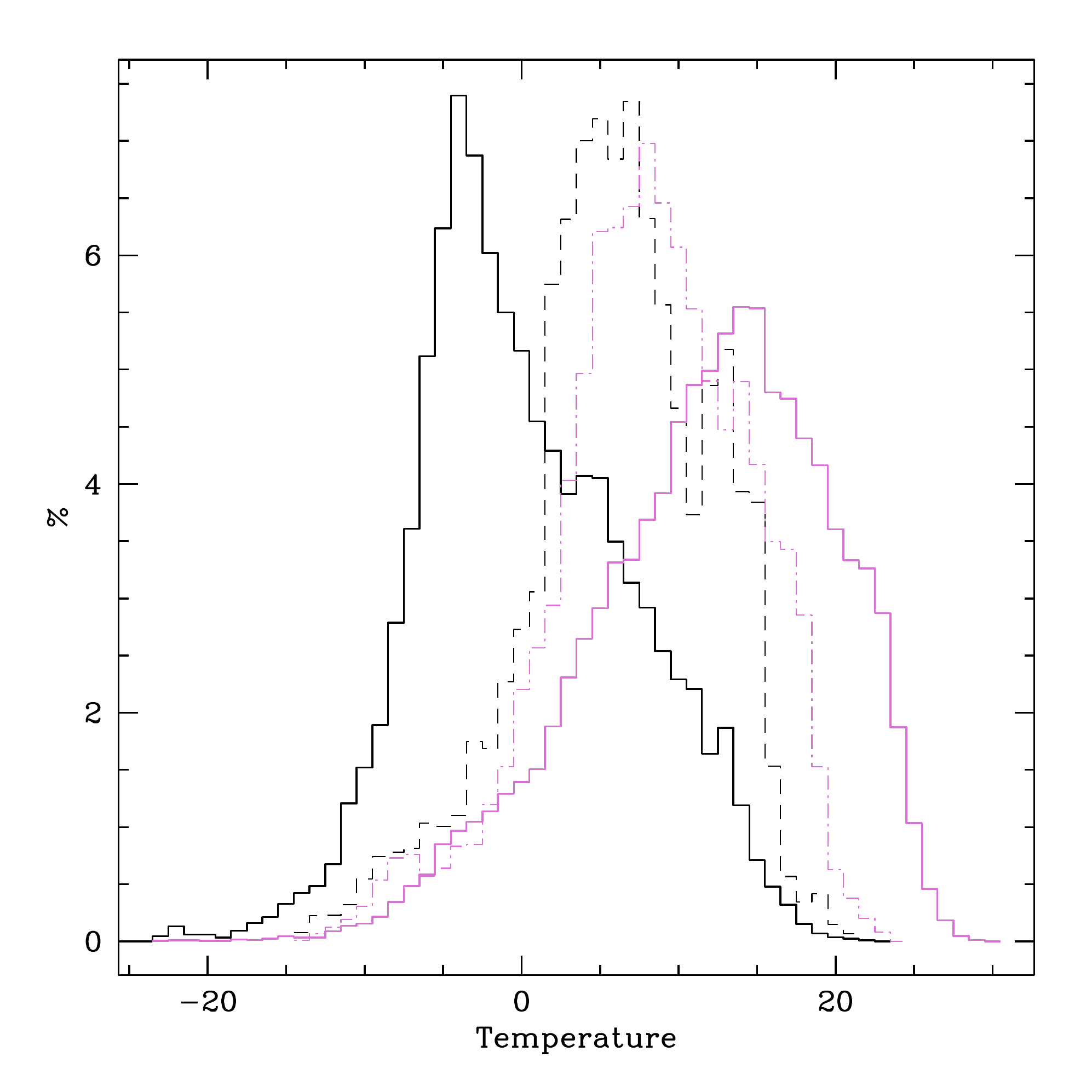}%
   \hfill
   \includegraphics[width=7cm,height=7cm]{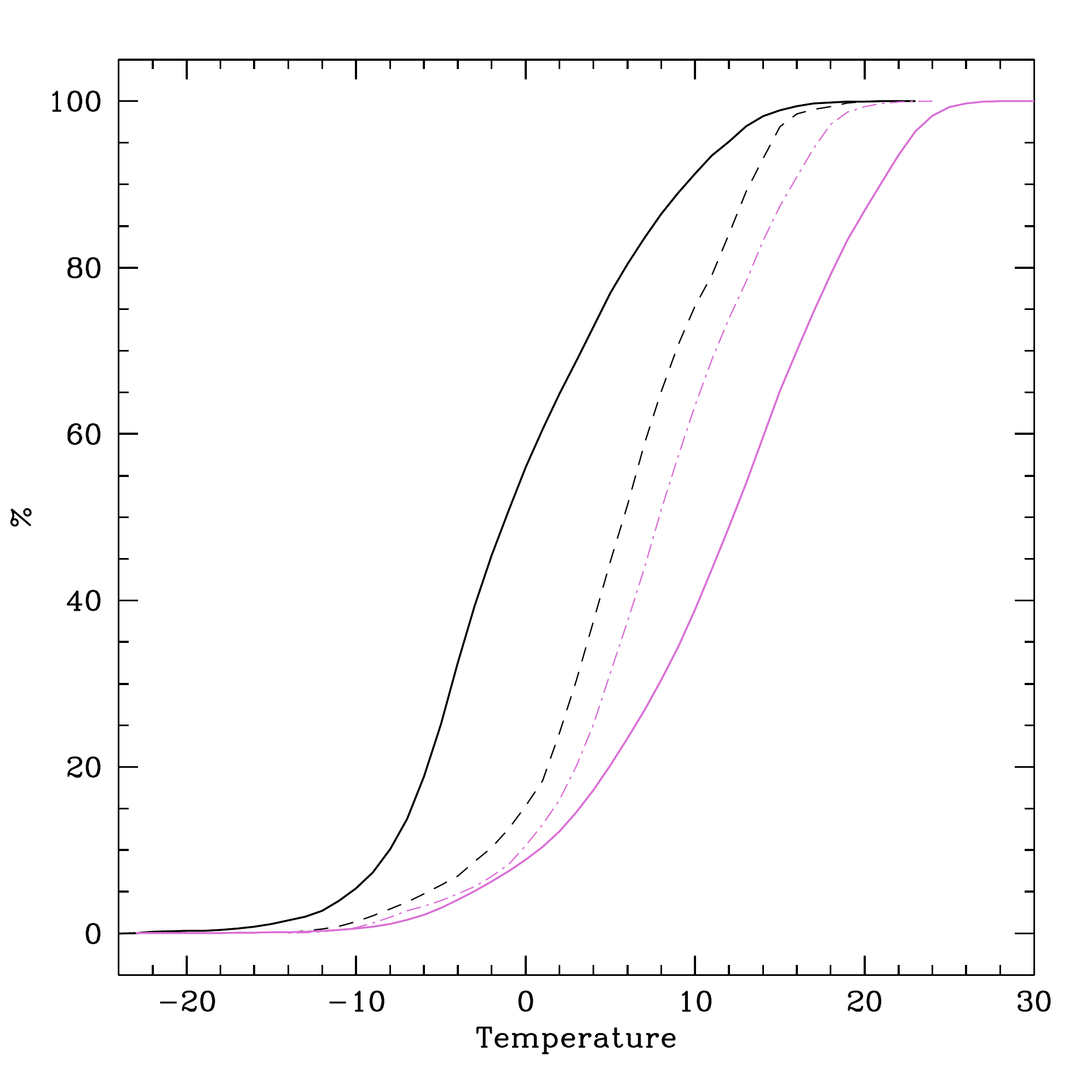}
   \caption{ Top: the temperature distribution at  Vallecitos and OAN. Bottom: the cumulative distribution function of the temperature. The color and line style demarcation is similar to Figure\,\ref{fig:widefig1}; the solid lines are reserved for  Vallecitos and dashed lines for  OAN. The nighttime is plotted in black. }
   \label{fig:temp}
\end{figure}

The wind patterns and velocity are distinct between the two stations, but are not significantly different. In contrast, the temperature and humidity measurements presented surprises. The temperature monitoring results are shown on the Figure \ref{fig:temp}.  At OAN, the temperature is  mild  most of the year with small differences between the night and the day.  In the left panel of Figure \ref{fig:temp}, the distribution of the daytime and nighttime  temperatures at OAN almost coincide throughout the year. Meanwhile, in the Vallecitos, they are well apart, with peaks  separated by 18$^{\mathrm o}$C. The temperature distribution in  Figure \ref{fig:temp}, as well as the cumulative  distribution, demonstrate that these two areas in close proximity have quite different temperature regimes, with the flat, lower-altitude Vallecitos having more extreme temperatures and larger gradients than the OAN. The low-temperature wing of the nighttime distribution, as well as the high-temperature wing of daytime curve, is steeper than the opposite sides, proving that extreme temperatures are less common and that usually the temperature stays in the reasonably mild  -17 to $+27^{\mathrm o}$C range. Nevertheless, temperature drops below $-20^o$C (with an absolute minimum of $-24^{\mathrm o}$C) have been registered three times for several  hours during these 18 months, which   represents 0.3\% of the monitored time.  Temperatures below $-20^{\mathrm o}$C violate the CTA temperature requirements.

\begin{figure}[!t]\centering
   \includegraphics[width=7cm, bb = 160 120 940 500, clip]{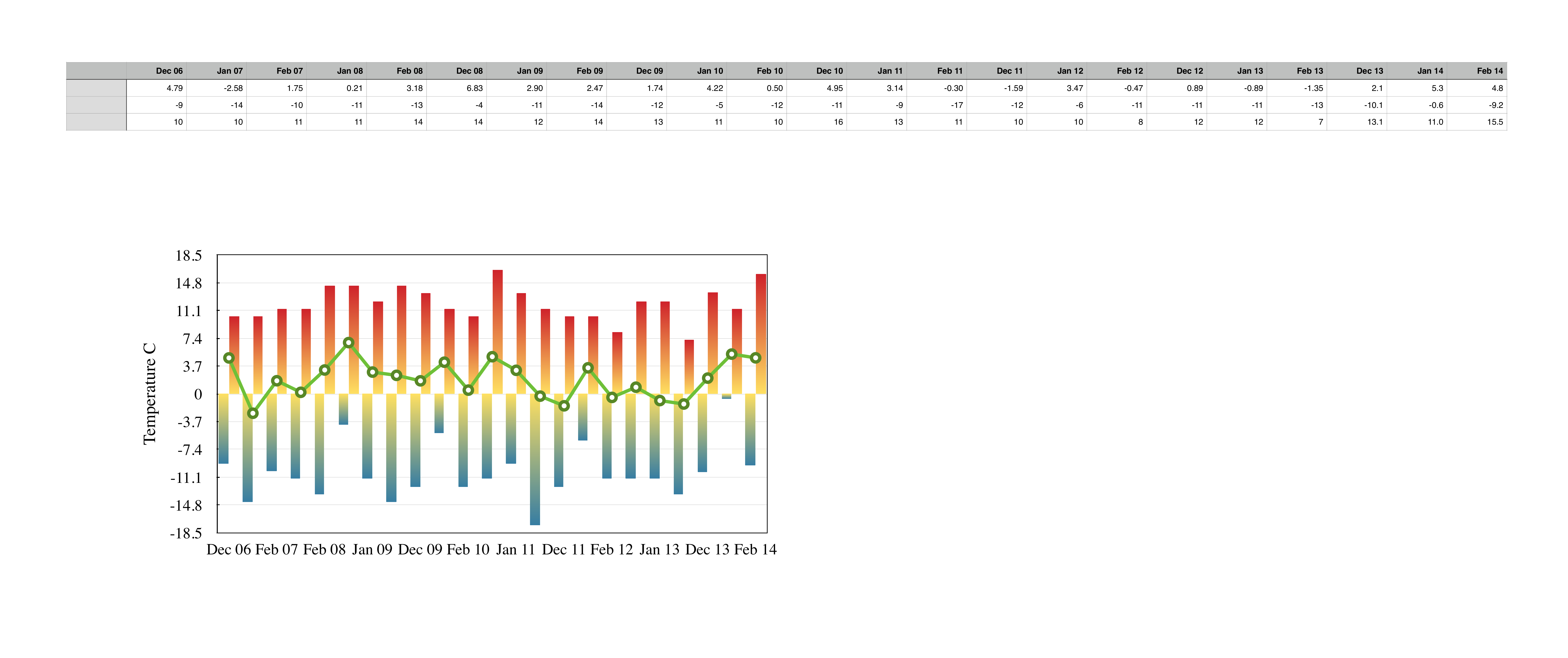}%
	\caption{Winter (December, January, February) temperature statistics of 7 years at SPM. The maximum registered temperature during each month is shown as red bars, the minimum as blue bars, and the monthly average as green circles. In 2013, for two winter months, the  average temperature was below zero. }
   \label{fig:result}

\end{figure}

\begin{figure}[!t]\centering
   \includegraphics[width=7cm,height=7cm]{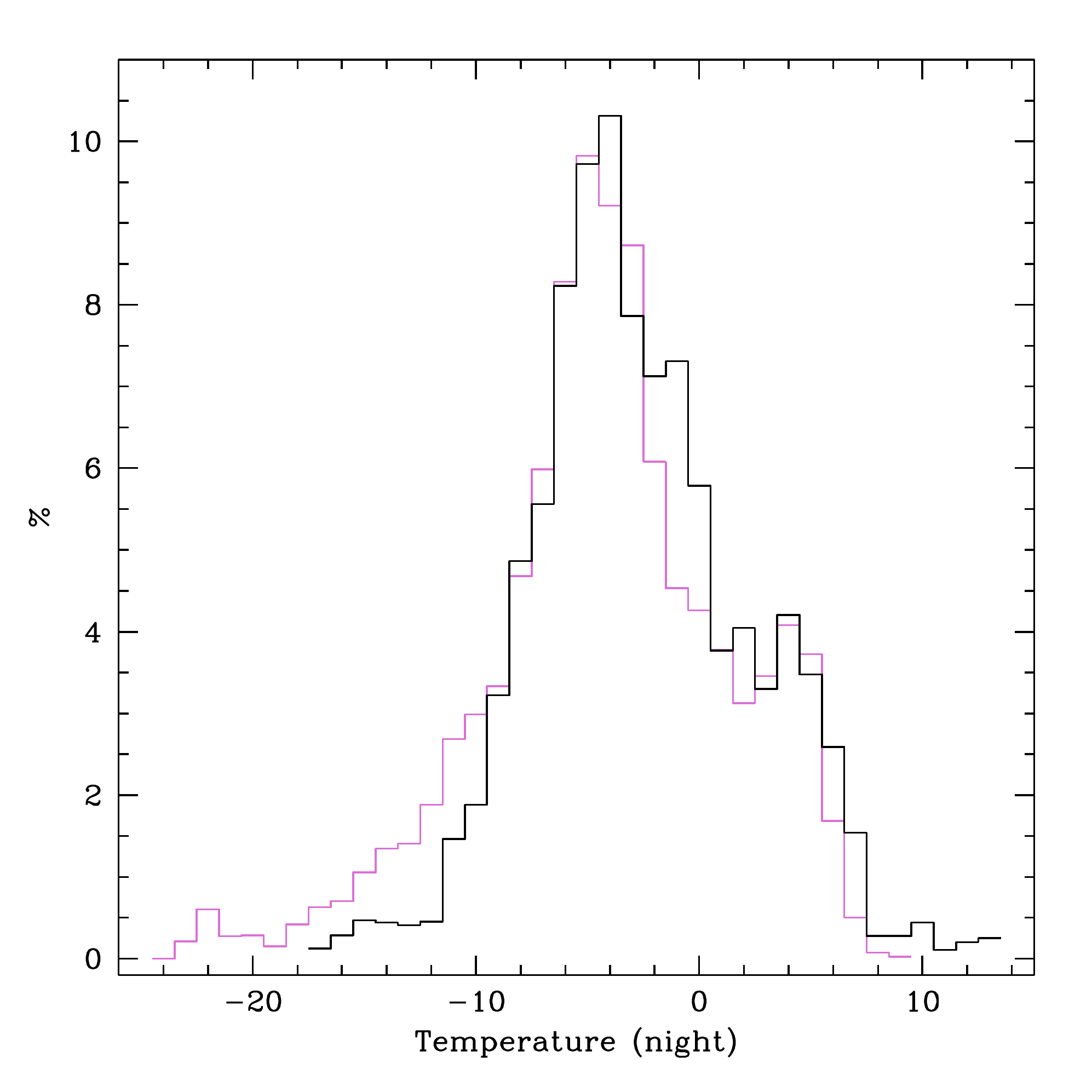}%
	\caption{Winter temperature distribution separately for two  years of monitoring. The first winter is from 2012 December to 2013 February (magenta line), and the second winter belongs to 2013 December to 2014 February (black line).}
   \label{fig:winter}
\end{figure}

Since no record exists of temperatures in Vallecitos prior to our study, we checked the monthly average temperature at SPM in three winter months during the last 7 years. Results in  the form of the maximum, minimum, and average temperature each month are presented in a graphical form in Figure\,\ref{fig:result}.  Out of a total 23 months, only in 6 was the average temperature was below zero, and two out of six has happened in winter 2013. So the average temperature at SPM reflects  extremes taking place in Vallecitos. We view this as an indication that the occurrence of such low temperatures is not frequent and may happen every once or twice in 7 years. For the   Vallecitos site,  the nighttime temperature  distributions of two winters (December, January, February) on record are presented  in  Figure \ref{fig:winter}.  All instances of extremely low  temperatures occurred  in 2012 to 2013.   In contrast, during the winter of 2013 to 2014, the temperature  never reaches $-20^{\mathrm o}$C and, in fact, stays above  $-17^{\mathrm o}$C at all times\footnote{While this paper was under review, we already have data for winter 2014/2015 with no occurrences of temperatures below $-20^{\mathrm o}$C.} 

\subsection{Humidity}

The precipitable water  vapor (or the depth of water in a column of the atmosphere if all the water in that column was precipitated as rain) is a manner of measuring water content in the air and has an exponential dependency on the altitude. 
The results of  the study by \citet{2010PASP..122..470O}, spanning a period of 4 years from early 2004 until the end of 2007, confirmed  that  the OAN site is dry and exceptionally good for astronomy research. The  data analysis shows that during winter, even though the SPM  site is about 230 m lower in elevation than Armazones, it is drier than the Armazones site. In general, astronomers prefer dry air, which can be important at an infrared wavelength range for spectral transparency. However, for the atmospheric Cherenkov observations, water content in the air is not essential, unless it is so high that it condensates on the optical components of the telescopes. The ATMOSCOPE has been monitoring the relative humidity to study conditions at the candidate sites.  Relative humidity is the ratio of the partial pressure of water vapor in an air--water mixture to the saturated vapor pressure of water at a given temperature.

\begin{figure}[!t]\centering
   \includegraphics[width=7cm,height=7cm]{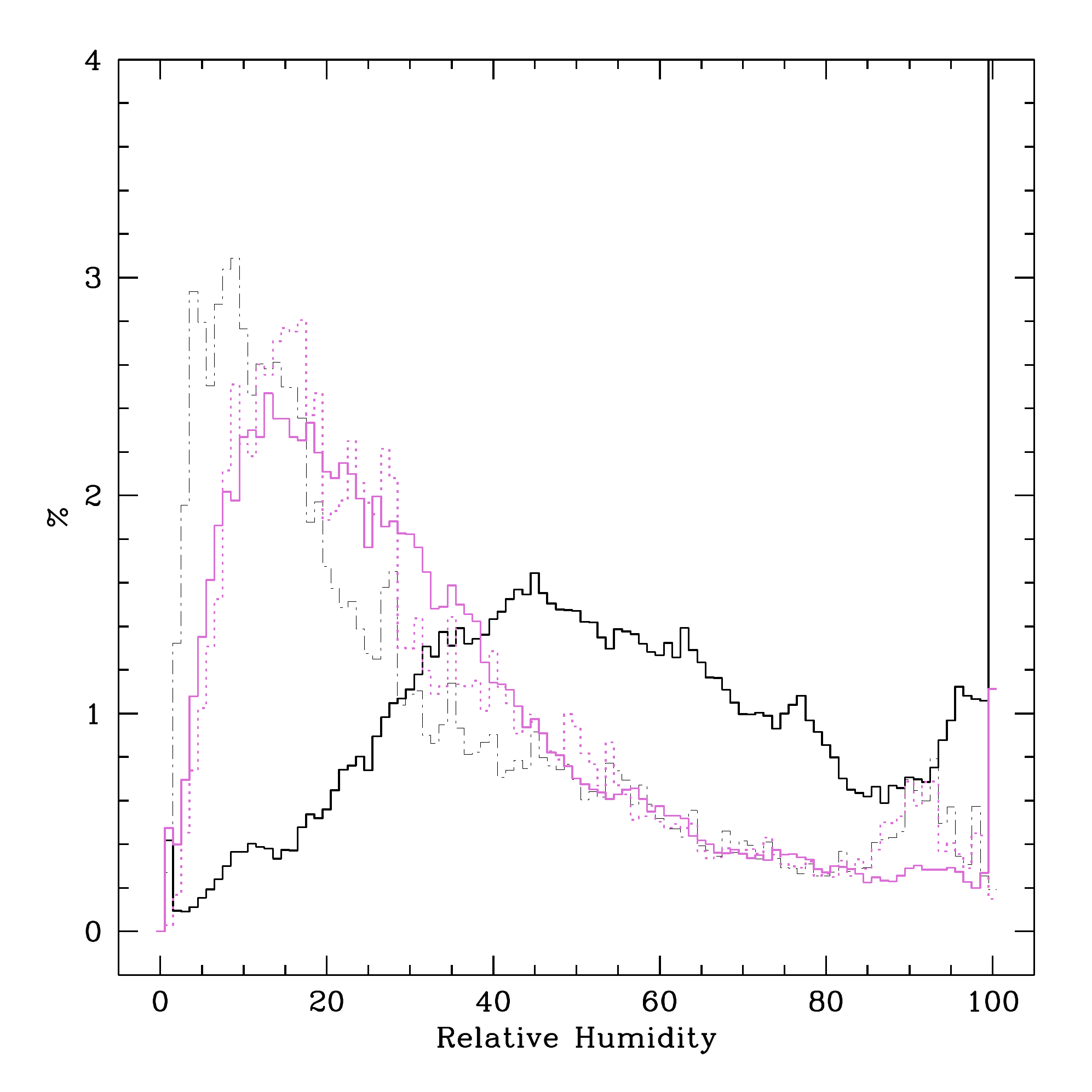}%
   \hfill
   \includegraphics[width=7cm,height=7cm]{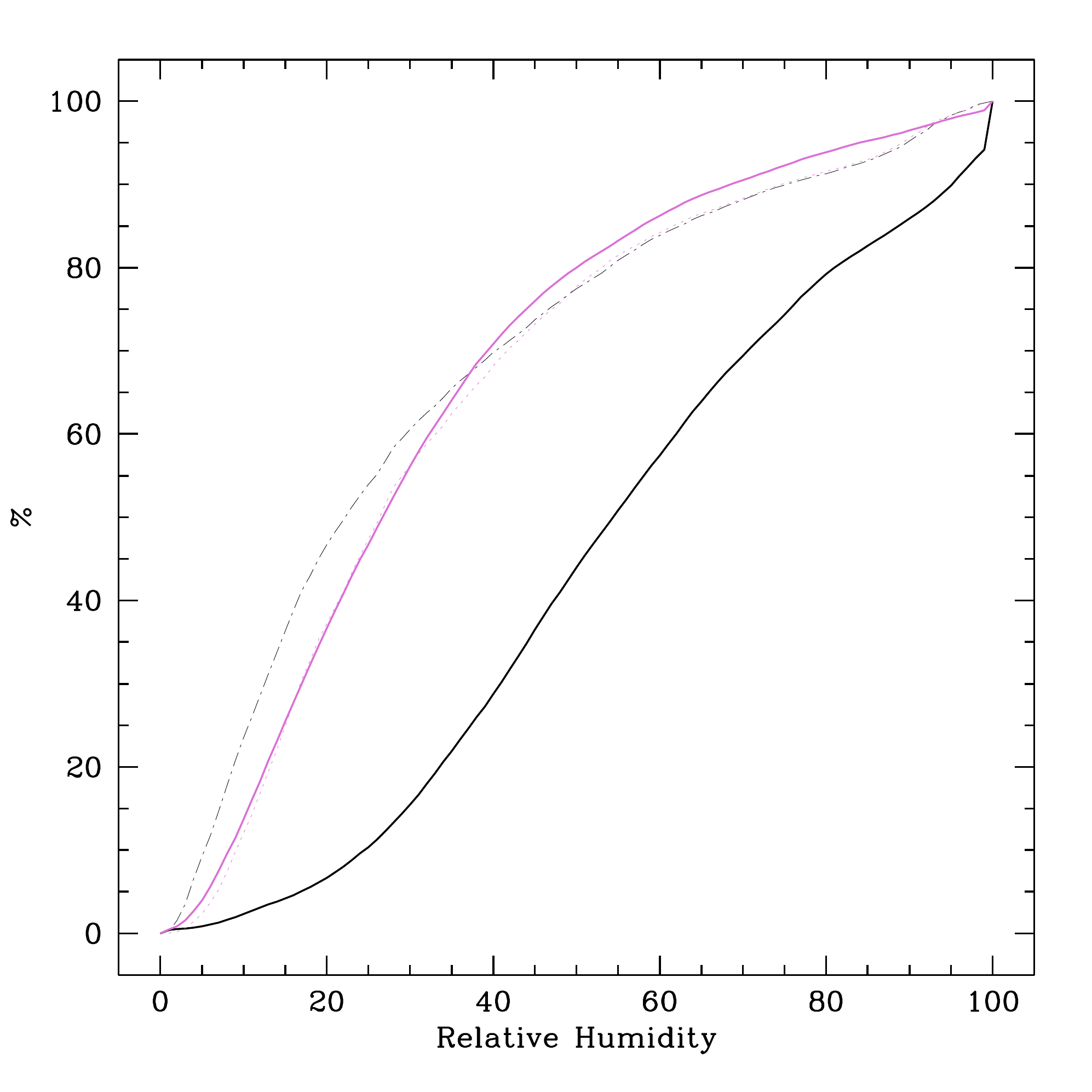}
   \caption{Top: the  humidity distribution histogram. Bottom: the cumulative distribution function of the humidity. The color and line style demarcation is similar to Figure\,\ref{fig:widefig1}, the solid lines reserved for  Vallecitos and dashed lines for  OAN. The nighttime is plotted in black. }
   \label{fig:humidity}
\end{figure}

As in the previous figure, in  Figure \ref{fig:humidity}, the relative humidity distributions are displayed in the form of histograms and cumulative distribution functions separately for daytime and nighttime for  Vallecitos and OAN.  At  OAN, the humidity is low around the clock, the nighttime being slightly drier than the day.  At Vallecitos, the daytime relative humidity is similar to OAN; in fact, it is less humid. However,  at night, humidity is outrageously high, being totally different from the OAN. The humidity distribution peaks around 40\%--60\%, dropping gradually toward 85\% only to produce a secondary peak at higher values.  During an unexpected  5.8\% of the monitored period of time, the humidity reaches 100\%.  Relative humidity is another parameter that we found strangely off of the corresponding values at the OAN.  High humidity in itself is not a problem for observations of Cherenkov radiation if the sky is clear. The problem is condensation of water on the mirror surfaces, which occurs when certain conditions are met.  In order to understand how much the high humidity at Vallecitos may affect the operation of CTA, we must derive the time losses due to water condensation on the telescopes.  For this, we can use the actual temperature and relative humidity and calculate the dew point.

The dew point is a water-to-air saturation temperature, i.e., the temperature at which the water vapor in the air at constant barometric pressure condenses into liquid water.  The dew point approaches the current temperature as the relative humidity  increases. 
 We calculated the dew point $T_{\mathrm {dp}}$ for Vallecitos and SPM according to the equations $\gamma= \ln \left( \frac{H_{\mathrm R}}{100} \right)+\frac{b \cdot T}{c+T}$ and $T_{\mathrm {dp}}=\frac{c \cdot \gamma}{b-\gamma}$, where $H_{\mathrm R}$ is the relative humidity, $T$ is the temperature, and  $b=17.67$, $c=243.5 ^{\mathrm o}C$ are constants \citep{1980MWRv..108.1046B}. When the dew point is equal to the current temperature,  the water starts to dew over solid surfaces. That is what  caused part of the problematic ASC images. According to calculations presented in  Table\,\ref{tab:hrlost}  during 17\% of the monitored time, the dew point was within 2$^o$C from the ambient temperature at Vallecitos, as opposed to only 7\% of the time at OAN. Those 7\%  probably correspond to the rain in the entire area, which is discounted from the observable time anyways. Part of the remaining 10\%  (about half) corresponds to cloudy weather and is also counted as not usable. However, there still remain a few percent (we do not know the exact number) when the sky might be clear, the humidity at the OAN low, but still, in Vallecitos,  water condensation may prevent normal operation of the observatory.   The errors of dew point determination by using cited empirical relations are of the order of 0.2$^{\mathrm o}$C, but we have chosen an order higher than  2$^{\mathrm o}$C difference of dew point temperatures from the ambient to make absolutely sure that time lost to the condensations on the surface of mirrors would not represent a problem at a lower heights (see below in  this section).   Worth mentioning is that in the total amount of monitored time (since it is not two full years), a fraction of the winter months is high, exacerbating the problem. 

\begin{figure*}[!t]\centering
   \includegraphics[width=7.45cm, bb= 10 0 560 560, clip]{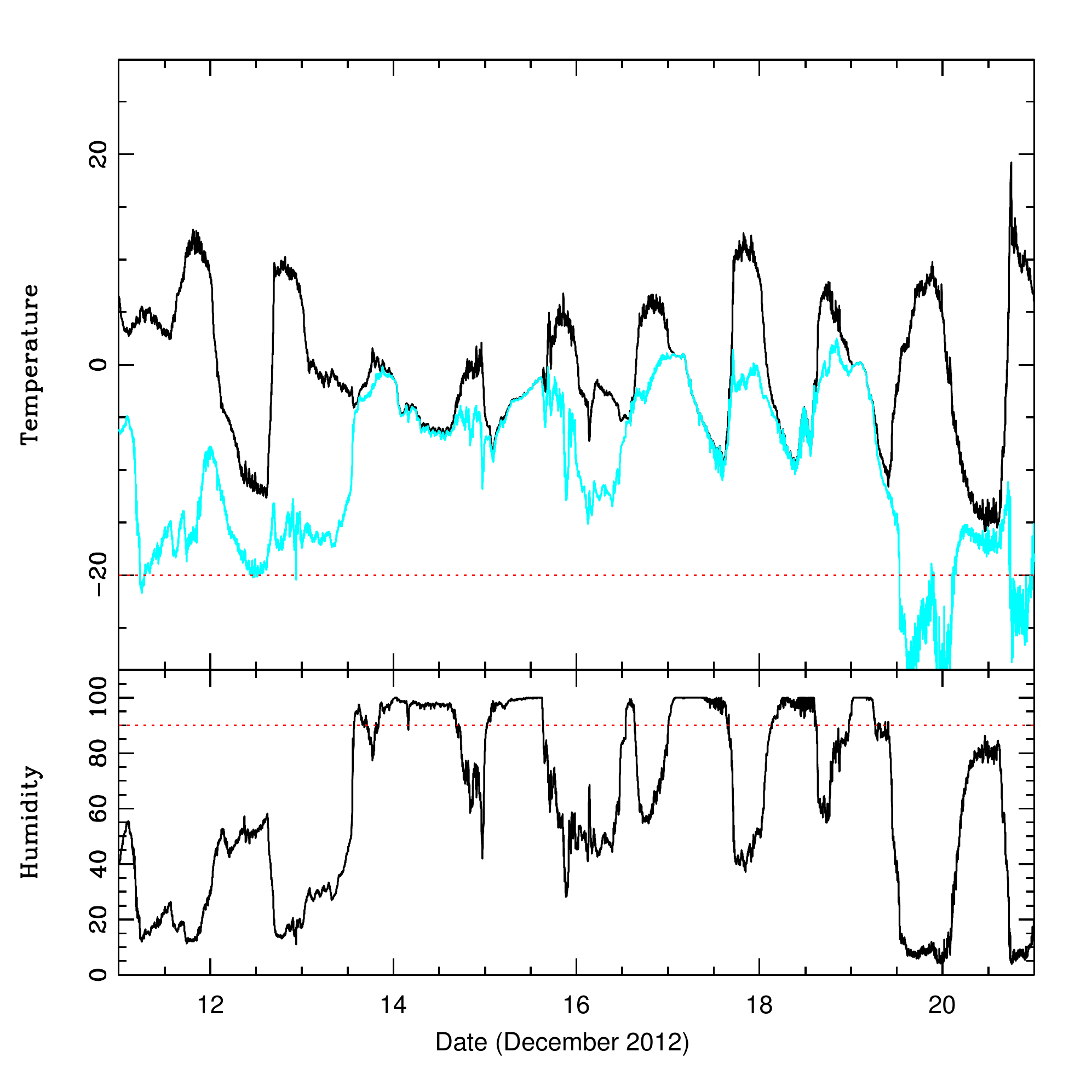}%
   \hfill
   \includegraphics[width=7cm, bb= 45 0 560 560, clip]{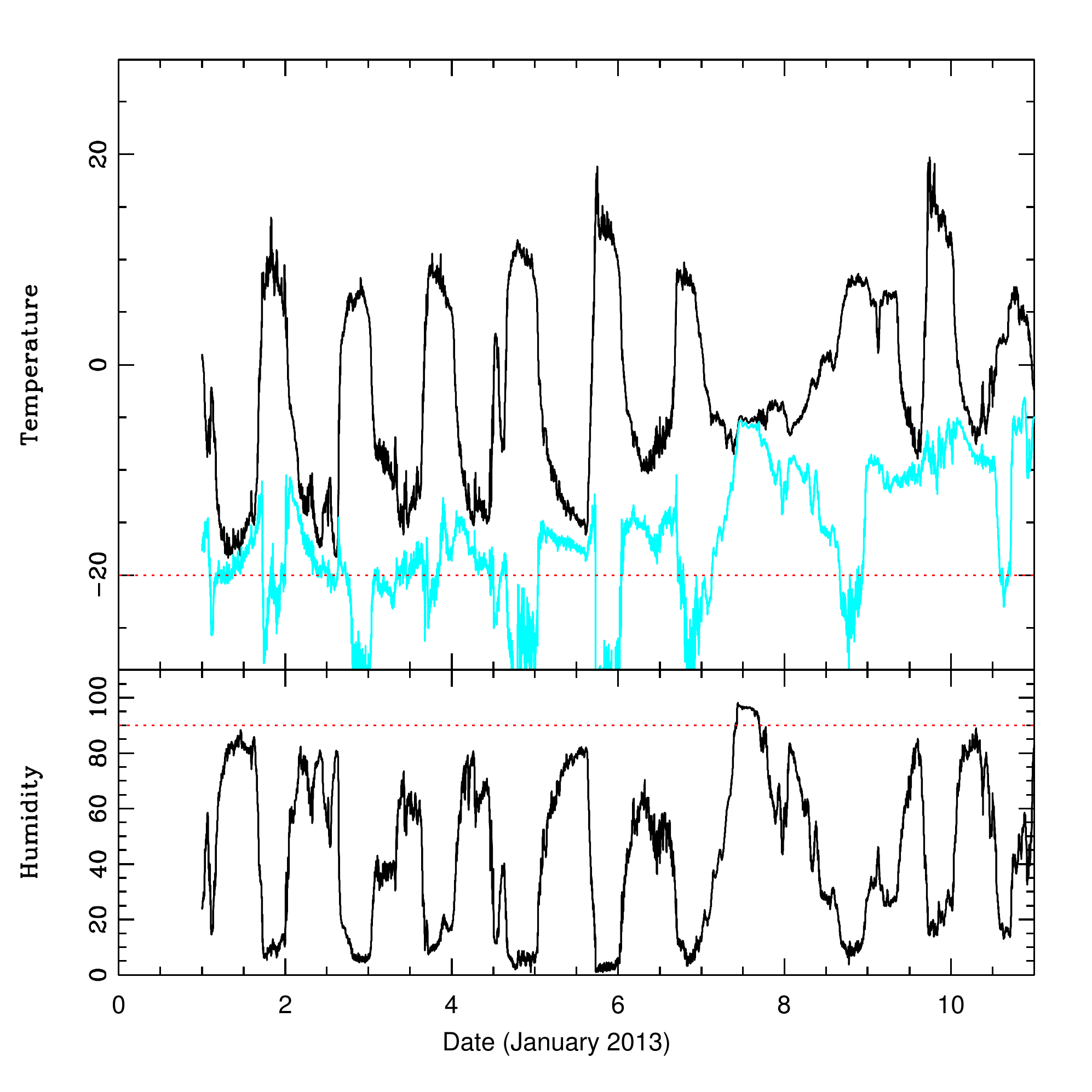}
 \caption{Two examples of temperature and humidity behavior  in winter months.  In the upper panel of each graph, the black line marks the measured temperature in Vallecitos, the cyan line corresponds to calculated dew point  ($T_{\mathrm {dp}}$), and the red dashed line indicates $-20^o$C.  In the bottom panel, we present the measurements of humidity at Vallecitos; the red dashed line indicates  the 90\% threshold.  }
   \label{fig:widefig7}
\end{figure*}

Examples of  interrelation between temperature, humidity, and dew point are shown in  Figure\,\ref{fig:widefig7}.  ATMOSCOPE measurements of temperature and humidity are plotted  for a  dozen days for two periods in winter 2012 to 2013. Overplotted is the dew point. In 2012 December,  there was a string of rainy days, and the plot  shows relative humidity exceeding 95\%  when the dew point was equal to the ambient temperature. In
2013 January,  the humidity rose every night as usual for  Vallecitos, but the dew point stayed lower than the air temperature, thus humidity was high, but did not reach the condensation point except for on one occasion.  

\begin{table}[!t]\centering
\caption{Lost time by T-2 $<$ T$_{dp}$}
\begin{tabular}{lrrr}
	\hline
	\hline
	Site  & Total  (hr) & Lost (hr) & Lost (\%) \\
	\hline
	\hline \\
	Vallecitos & $6010$ &	$1042$ &	$17.3$ \\
	SPM & $5865$ & $430$ & $7.3$ \\
	\hline
	  \hline
\end{tabular}
 \label{tab:hrlost}
\end{table}

%

\begin{figure}[!t]\centering
\includegraphics[width=7cm,height=7cm]{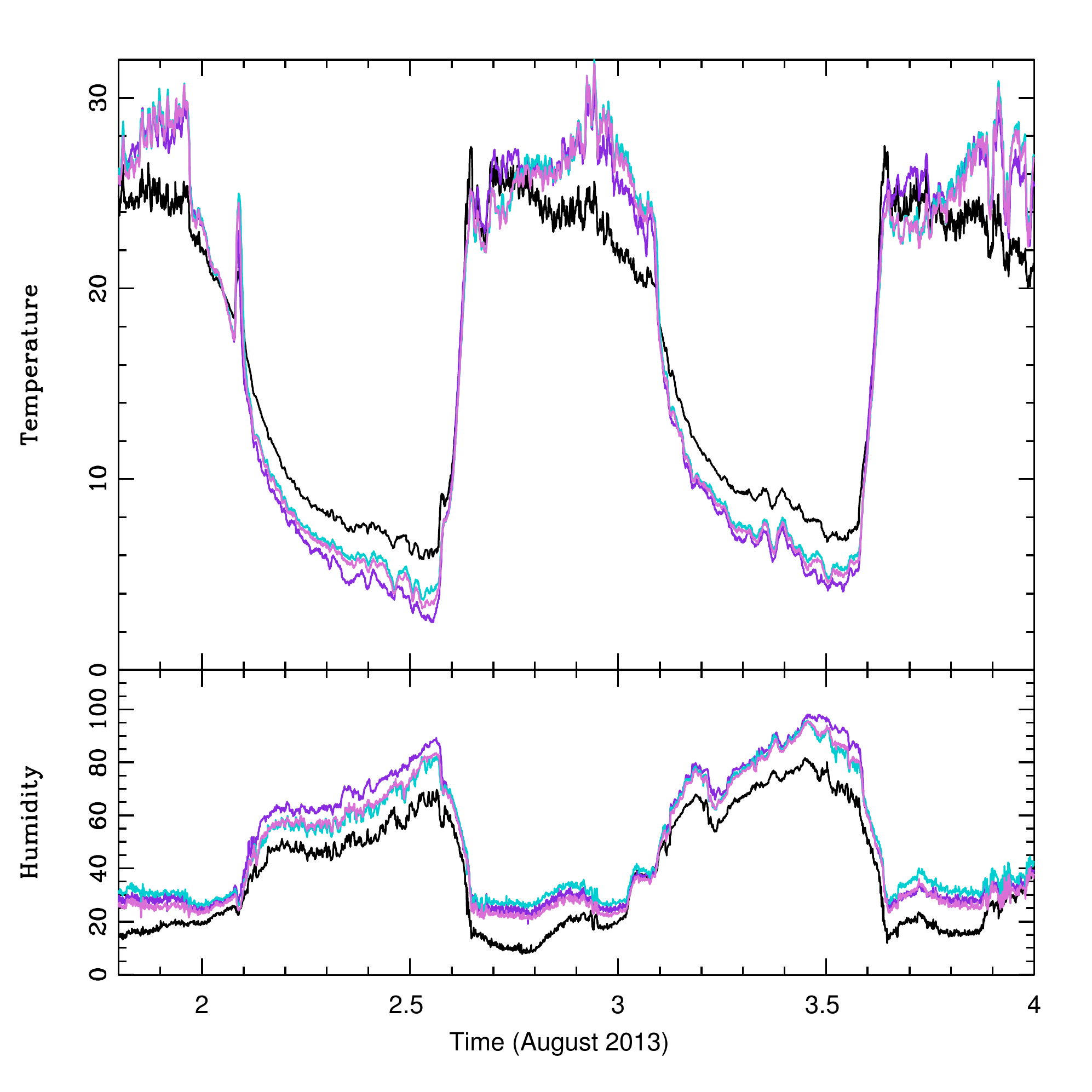}
   \caption{ Another example of temperature and humidity monitoring. The plot  demonstrates clear dependence on temperature and humidity from the height above the ground.  The colors are assigned as follows:  black, 10\,m (ATMOSCOPE); cyan, 7\,m; dark magenta, 3\,m; violet, 1\,m. During the night, the  temperature falls, gradually decreasing during the night. Closer to the ground, the lower is the temperature and higher is the humidity. The trend reverses during the day.}
   \label{fig:fig10}
\end{figure}

Another interesting observation is that the temperature and humidity at Vallecitos are very sensitive to the height from the ground.   We operate here with the data taken by the  ATMOSCOPE  at 10m above the ground,  but additional humidity and temperature sensors at different elevation from the ground (1, 4 and 7 m) indicate that the temperature falls drastically closer to the ground after sunset and warms up higher than the air temperature at 10 m height during a sunny day. Meanwhile the humidity in the valley after the sunset  increases closer to the ground than in the air. 
This behavior is displayed in Figure\,\ref{fig:fig10}. Although the sensors below ATMOSCOPE are not calibrated and are not  very precise, the figure leaves no doubt that closer to the ground, the lower is temperature and higher is the humidity. Also, it is apparent that the temperature decreases during the night reaching the lowest point right before the Sun rises; similarly, the humidity  increases gradually, reaching the highest value by the morning. This is an example of three days, but the pattern is very common throughout the year, particularly in cloudless time.  On average, a two-degree difference is observed between night temperatures at 10 and 3 m heights. 

The registered temperatures and humidity at Vallecitos should
 be taken into account in the telescope design. We believe that 
they can  influence the operation of the ACTs and might represent a technical challenge.

\section{Essential Differences between SPM and Vallecitos and Probable Reasons}
 \label{sec:conclu}

\begin{figure}[!t]\centering
\includegraphics[width=7cm,height=7cm]{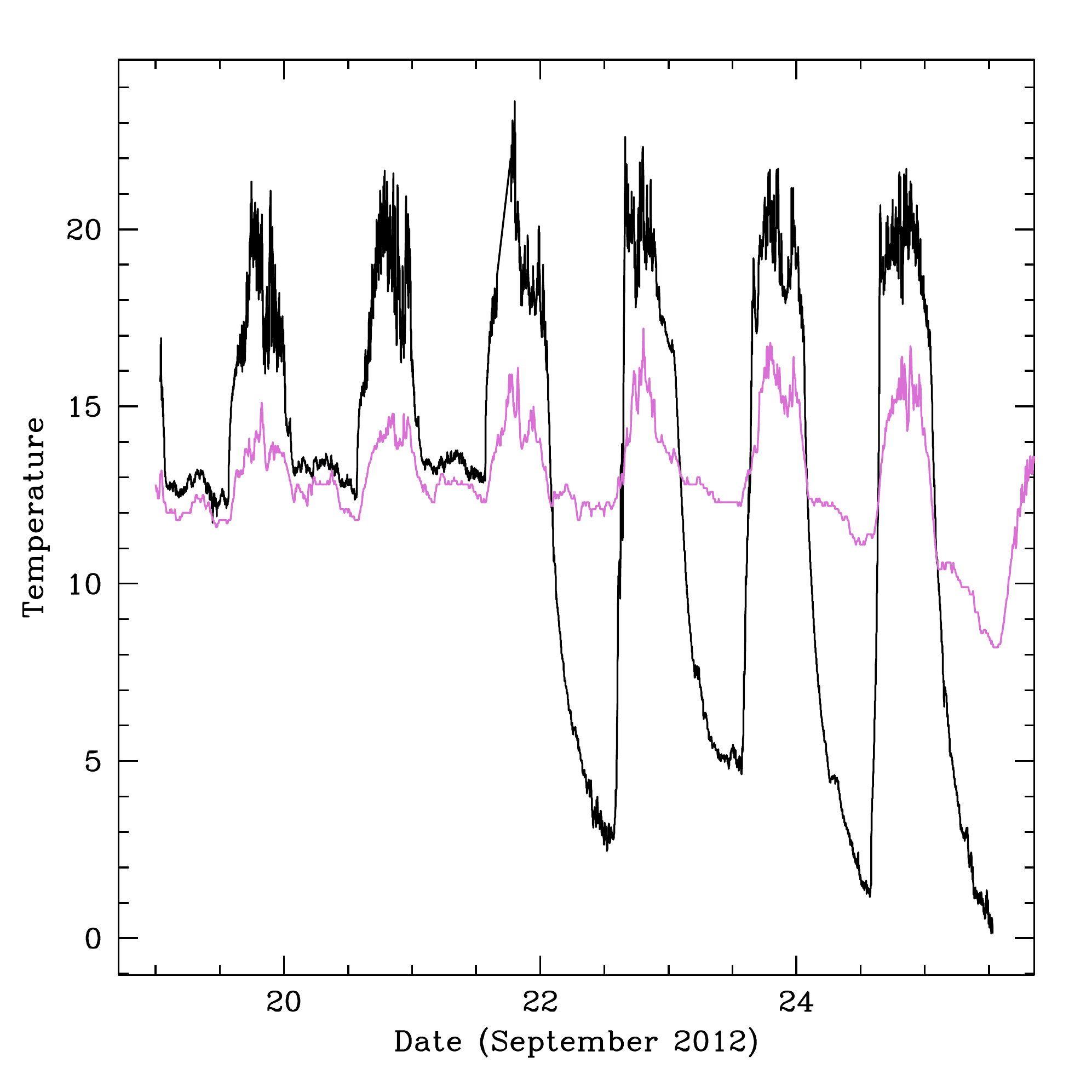}
   \caption{Temperature  variation depending on the location around SPM. The smallest fluctuation plotted in magenta corresponds to the OAN.
   The measurements by ATMOSCOPE are plotted in black; the instrument was installed at the observatory campus before transfer to the designated CTA area on  September 21.  The lower the altitude the larger the temperature gradient between daytime and nighttime.}
   \label{fig:fig11}
\end{figure}

 It turns out that at nights the  valley  is  a much colder place than the observatory  situated on the surrounding hills at a higher altitude. In fact, the temperature gradient increases with the decrease of altitude toward the valley. In Figure\,\ref{fig:fig11}, we present temperature measurements at  two distinct locations. The ATMOSCOPE was installed at the residential campus, half-way between the observatory weather station and Vallecitos, from 2012 June to September. It was transferred and installed at the center for the proposed site for CTA on 2012  September 21. The black line in Figure\,\ref{fig:fig11} is temperature measured by the ATMOSCOPE. It is notable that during the day it is much warmer at the  campus than by the OAN telescopes (the magenta line shows the temperature at the observatory). It is also  evident that at nights the temperature is much lower (about 10$^{\mathrm o}$C) at  Vallecitos than at the campus or the observatory.  This large difference is observed regularly, except for times when the entire area is covered by low clouds. At those times, the temperature difference is less marked. The relative humidity at night also behaves differently than at the observatory, the average humidity being often in 40-60\% range with high percentage of occurrences when the humidity is above 90\%.  Comparison of a two-year record between the two sites shows that in Vallecitos the dew point reaches 10\%  more often than at the observatory. The humidity at night at Vallecitos is higher with temperatures lower, closer to the ground than at 10\,m height. The soil study showed that there is no underground water down to 6\,m, and the solid granite starts at $\approx2$\,m depth.

\begin{figure*}[!t]\centering
\includegraphics[width=14cm,height=8cm]{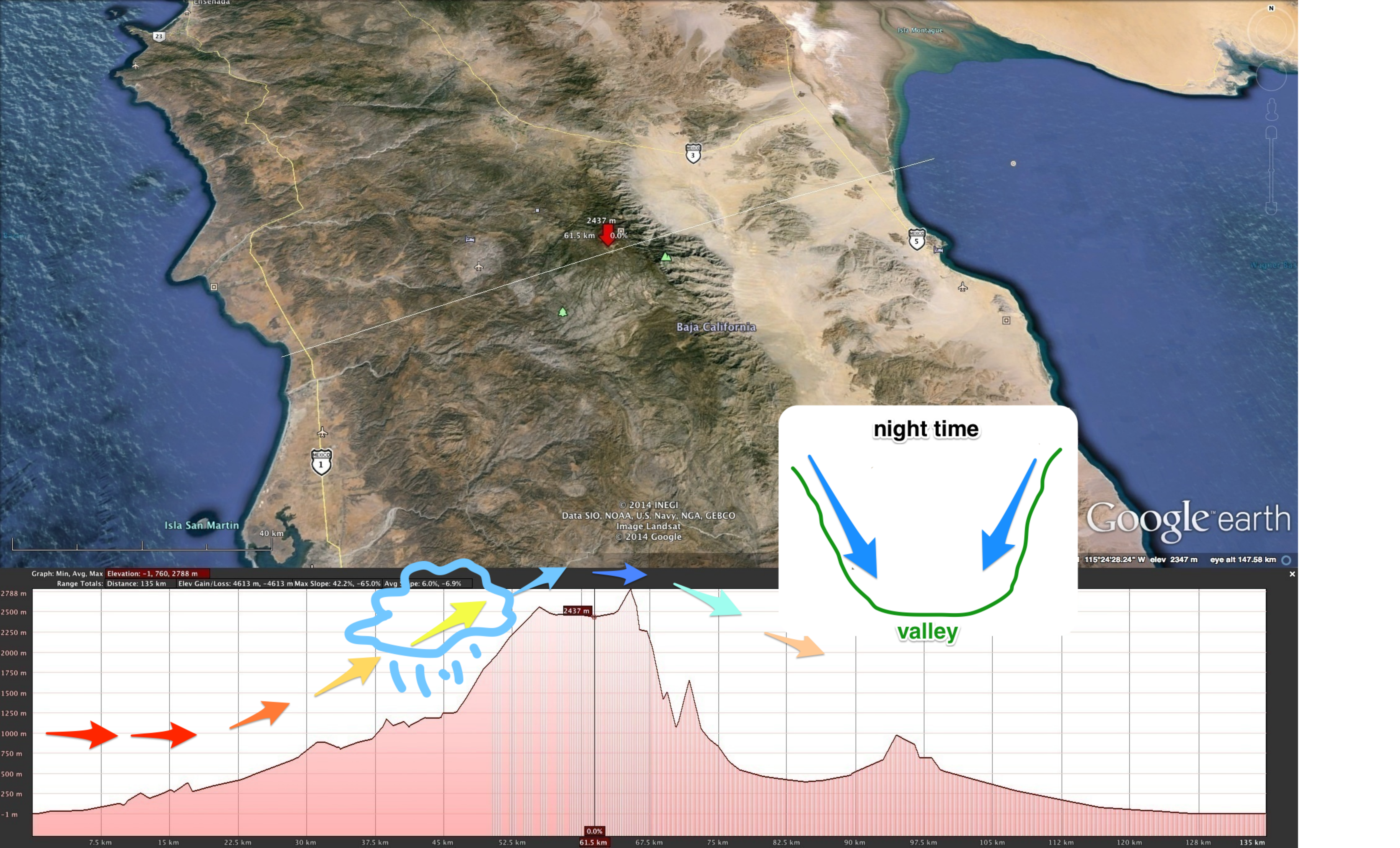}
   \caption{ Cut across the peninsula along the direction of the prevailing wind shows a profile typical for orographic precipitation,  when warm, moist air is lifted as it moves over a mountain range. As the air rises and cools, orographic clouds form and serve as the source of the precipitation, most of which falls upwind of the mountain ridge. Meanwhile, in the valley at the center of the mountain range, a katabatic wind during the night cools the bottom of the valley, while during the day it warms up by  anabatic wind (breeze).}
   \label{fig:fig12}
\end{figure*}

We speculate that such a temperature/humidity regime may result from orography. The prevailing air flow from the Pacific creates a dome over the central bowl-like part of the SPM mountain chain, trapping the air and humidity as if in  a greenhouse. Meanwhile, the local air day/night circulation is established in the central valley. During the sunny daytime, Sun-warmed air gently flows up the mountain slopes, gradually cools at higher altitudes, and then descends back to the central part of the valley.  At night, the  circulation of air reverses. The cold, dense air is drained down the slopes  via the katabatic wind and accumulate at the bottom of the valley in a pool of cold air  (Figure\,\ref{fig:fig12}).  Meanwhile, the slopes remained relatively warm. There might be other reasons contributing to the peculiar weather in  Vallecitos that we failed to recognize.  However, it can not be ignored that the presence of this low valley filled with cool air next to the observatory  contributes to the excellent atmospheric conditions of the OAN at night.  The registered temperatures and humidity at Vallecitos should not influence  the  operation of the ACTs,  but might represent  a technical challenge. 

\section{Conclusions}
 \label{sec:conclu}

We conducted an extensive study of weather conditions in an area proposed to host CTA-north array at SPM, M\'exico. The area in question is a flat, treeless valley, $\approx400$\,m below the optical observatory located there. We compare here records (historic) obtained at the observatory (OAN), including the weather station there, with those we collected in almost two years at the Vallecitos  using the ATMOSCOPE data.  Some global and important parameters as the fraction of cloudless nights and darkness of the sky are similar for the two locations at about 3.5 km from each other. Observations from Vallecitos confirmed a very high fraction (above 80\%) of clear nights and extremely low night sky background, making the area an exceptional site  for ground-based astronomical research. We also show that the  Vallecitos site is calm regarding wind speeds and is safe for ACTs installed without protection.  In the meantime, significant differences between the two locations have been observed regarding temperature and humidity extreme levels and gradients.  The temperatures as low as $-24^{\mathrm o}$C and as high as $+31^{\mathrm o}$C have been registered. The humidity is quite similar in SPM and Vallecitos during the day, but at night the humidity is high in Vallecitos, with 17\% of total time reaching within $2^{\mathrm o}$C of the dew point. 


SPM is confirmed to be one of a few sites around the planet with outstanding conditions suitable for astronomical research.

 We  acknowledge the support of Institute of Astronomy of UNAM for the maintenance and supplementary investment in  the ATMOSCOPE. M.S.H. is grateful  to the IA UNAM for the financial support during this research, which was partly covered through the CONACyT grant "Laboratorios Nacionales" No.260369. We would like to thank  all technical staff of the OAN, for their maintenance work on the ATMOSCOPE. G.T. has been supported by PAPIIT grant IN\,107712. The authors gratefully acknowledge the support by the projects LO1305, LG14019, and LE13012of the Ministry of Education, Youth and Sports of the Czech Republic. We appreciate insightful comments made by Manuel Alvarez, a pioneer who  explored the area.  We are also thankful to a broader CTA community who read and commented on the paper, particularly our Spanish colleagues who critically assessed the manuscript. We are grateful to Michael Prouza, CTA internal referee, for careful assessment of the manuscript.

\end{document}